\documentclass[fleqn,usenatbib]{mnras}

\usepackage{tabularx}
\usepackage{newtxtext,newtxmath}

\usepackage[T1]{fontenc}
\usepackage[integrals]{wasysym} %

\DeclareRobustCommand{\VAN}[3]{#2}
\let\VANthebibliography\thebibliography
\def\thebibliography{\DeclareRobustCommand{\VAN}[3]{##3}\VANthebibliography}

\usepackage{graphicx}	
\usepackage{amsmath}	
\usepackage{fix-cm}
\usepackage{makecell}  

\usepackage{lineno}

\usepackage{lineno}

\title[Radio Emission from the K2-18 System]{Upper Limits on Radio Emission from the K2-18 System}

\author[Kelvin Wandia et al.]{
Kelvin Wandia,$^{1}$\thanks{E-mail: kelvin.wandia@manchester.ac.uk}
Chenoa Tremblay ,$^{2,3,4}$
Michael A. Garrett, $^{1,5}$
Alex Andersson, $^{6,7}$ 
Megan G. Li, $^{8}$
\newauthor
Vishal Gajjar, $^{2,3}$
Robert J. Beswick, $^{1}$
Jack F. Radcliffe, $^{1,9}$
David R. DeBoer, $^{3,6,7}$
P.B. Demorest, $^{10}$
\newauthor
Daniel Czech, $^{6,7}$
Wael Farah$, ^{2,3}$
Ian Heywood, ${^7}$
Andrew Siemion, $^{1,3,6,7,11}$
\\
$^{1}$Jodrell Bank Centre for Astrophysics, University of Manchester, M13 9PL, UK\\
$^{2}$SETI Institute, 339 Bernardo Ave, Suite 200, Mountain View, CA 94043, USA\\
$^{3}$Berkeley SETI Research Center, University of California, Berkeley, CA 94720, USA \\
$^{4}$Department of Physics and Astronomy, University of New Mexico, Albuquerque, NM 87131, USA\\
$^{5}$Leiden Observatory, Leiden University, PO Box 9513, 2300 RA Leiden, The Netherlands\\
$^{6}$Breakthrough Listen, Astrophysics, Department of Physics, The University of Oxford, Keble Road, Oxford, OX1 3RH, UK \\
$^{7}$Astrophysics, Department of Physics, University of Oxford, Keble Road, Oxford, OX1 3RH, UK \\
$^{8}$Department of Earth, Planetary, and Space Sciences, University of California, Los Angeles, CA 90095, USA \\
$^{9}$Department of Physics, University of Pretoria, Lynnwood Road, Hatfield, Pretoria, 0083, South Africa \\
$^{10}$National Radio Astronomy Observatory, P.O. Box O, Socorro, NM 87801, USA\\
$^{11}$University of Malta, Institute of Space Sciences and Astronomy, Msida, MSD2080, Malta \\
}

\date{Accepted XXX. Received YYY; in original form ZZZ}

\pubyear{\the\year{}}

\begin{document}
\label{firstpage}
\pagerange{\pageref{firstpage}--\pageref{lastpage}}
\maketitle
%
\begin{abstract}

Stellar and planetary magnetic fields play a crucial role in the habitability of a planet and the integrity of its atmosphere. The detection of methane and carbon dioxide, along with a tentative identification of the potential biosignature  dimethyl sulfide/disulfide, in the atmosphere of K2-18 b, a sub-Neptune orbiting an M dwarf star present an intriguing question regarding the stellar magnetic environment and the resistance of the planet's magnetosphere (if it exists) to erosion by magnetic activity from the host. To probe for radio emission from the system, we have conducted observations using the Karl G. Jansky Very Large Array (VLA) at S, C and X-bands (2-4, 5.5-7.5 and 8-10 GHz respectively) to search for coherent and incoherent radio emission. We detect no radio emission associated with incoherent emission mechanisms. We report $3\sigma$ Stokes I upper limits of $49.8\ \mu\rm{Jybeam}^{-1}$ at S-band, $17.7\ \mu\rm{Jybeam}^{-1}$at C-band and $18.0\ \mu\rm{Jybeam}^{-1}$  at X-band and an upper limit of the ratio of the radio to the total bolometric luminosity of $\log L_\text{R}/\log L_\text{bol}<-8.8$.  We have also searched for short duration bursts associated with coherent emission mechanisms at C and X-bands . No signals above a $3\sigma$ significance threshold are detected. Although no signals are detected our radio observations offer constraints, albeit limited, on the stellar magnetic environment supporting recent X-ray observations indicating K2-18 is a very faint emitter. Our results also contextualise any planetary transmission spectra by providing constraints on the activity level of the host.

\end{abstract}


\section{Introduction}
\label{sec:intro}

The solar neighbourhood is dominated by low mass, low luminosity stars of spectral type M commonly referred to as M dwarfs. This population has been the focus of comprehensive studies that have reported prevalence rates as high as $70-75\ \%$ of the total main sequence stellar population in the solar neighbourhood \cite[][]{Henry2006,Winters2019}. M dwarfs have masses of $0.075-0.61\ \rm{\textit{M}_{\odot}}$ \citep[][]{Todd2024} and effective temperatures of $2500-4000$ K \citep{Mamajeck2013}. \textcolor{black}{Due to their prevalence and low masses, M dwarfs are favourable targets for radial velocity surveys for exoplanet detection, as they exhibit larger stellar reflex motions compared to stars of other spectral types. The low luminosity of M dwarfs also places the habitable zone close to the star, resulting in short orbital periods \citep[][]{Shields2016}. These factors result in a large observational signature for hosted planets 
making them well-suited for transmission spectroscopy aimed at characterising exoplanetary atmospheres \citep[][]{Kaltnegger2011,Cloutier2017}.}
 The internal structure of M dwarfs transitions at masses of $0.31-0.34\rm\ {\textit{M}_{\odot}}$ \citep{Baraffe2018} from a radiative core surrounded by a convective envelope akin to our Sun to a fully convective interior. M dwarfs whose mass fall above the transition limit are believed to generate and sustain magnetic fields through an $\alpha\Omega$ dynamo \citep[][]{Parker1955} and for those \textcolor{black}{below} the limit, the exact mechanism remains uncertain although an $\alpha^2$ dynamo has been proposed \citep[][]{Chabrier2006,Dobler2006,Browning2008}. These dynamos are driven by stellar rotation and convection \citep[][]{Charbonneau2010,Reiners2014}.  Although magnetic fields are prevalent in all stars \citep[][]{Mathys2001,Berdyugina2009},  M dwarfs exhibit enhanced magnetic fields compared to main sequence stars of earlier spectral types \citep[][]{Shulyak2019} due to large convective eddies in their deep (or fully) convective zones, which lead to large-scale organised magnetic fields \citep[][]{Chabrier2006}.

The strong internal magnetic fields drive significant surface activity in the form of stellar spots, plages, faculae, and flares. \textcolor{black}{This activity results in surface inhomogeneities which is a  potential source of bias for absorption features in transmission spectra \citep[][]{Barclay2021_inhomogeneous}.} The magnetic fields also heat the stellar atmosphere leading to chromospheric emission in the optical \citep[e.g.][]{Gunther2020}, ultraviolet \citep[e.g.][]{France2013} and coronal emission in the X ray \citep[e.g.][]{Caramazza2023} and radio \citep[e.g.][]{Callingham2021}.  At radio wavelengths, particularly at centimetre wavelengths, the dominant process responsible for incoherent emission from M dwarfs is the gyrosynchrotron radiation mechanism. Gyrosynchrotron radiation arises from mildly relativistic electrons in a hot plasma from thermal (the high-energy tail of a Maxwellian distribution) \citep[][]{Golay2023} or non-thermal energy distributions \citep[][]{Dulk1979,Dulk1985}. The radio emission observed from gyrosynchrotron processes occurs at harmonics $s\  (\text{typically} 10\lesssim s\lesssim100) $ of \textcolor{black}{the local cyclotron frequency $\nu_c\ \rm{(MHz)} = 2.8\times B\ \rm{(gauss)}$ } \citep[][]{Gudel2002}. Another process that occasionally contributes to incoherent emission is synchrotron radiation from highly relativistic electrons in a non-thermal energy distribution. For detected incoherent radio emission, the mechanism responsible can be determined using the degree of circular polarisation commonly quantified through measurements of the Stokes V parameter. Radio emission from gyrosynchrotron processes exhibits a \textcolor{black}{moderate} degree of circular polarisation, while emission from synchrotron processes exhibits negligible circular polarisation but is often highly linearly polarised. Finally, we note that the inverse Compton catastrophe limits the brightness temperatures of incoherent mechanisms to $\lesssim10^{12}$ K \citep[][]{Kellerman1969}.

In contrast, radio emission with brightness temperatures far in excess of  $10^{12}$ K have been detected from stars \citep[e.g.][]{Callingham2021} which is indicative of coherent emission mechanisms. Such emission has been attributed to plasma emission and electron cyclotron masers. In the former, energetic beams of electrons or other instabilities cause perturbations in the plasma which excite Langmuir waves at the local plasma frequency $\nu_\text{p}\sim9\sqrt{n_\text{e}}$ kHz. The Langmuir waves are then converted to radio waves due to non-linear wave interactions in the plasma \citep[][]{Reid2017}. Plasma emission is observed in the first and second harmonics \citep[][]{Melrose1980} and is dominant at frequencies < 1 GHz \citep[][]{Pohjolainen2020}. In contrast, the electron cyclotron maser emission \citep[ECME;][]{Wu1979} arises due to population inversion caused by electrons trapped in a flux tube leading to magnetic mirroring of electrons with large pitch angles. The distribution of the electrons is uncertain, although loss-cone and horseshoe distributions have been proposed \citep[see][]{Melrose2017}. The anisotropy in the electron distribution results in maser action with stimulated emission occurring at  $\nu_c$ \citep[][]{Melrose2017}. 

Radio emission generated by the mechanisms discussed above has been detected from M dwarfs at radio wavelengths. Quiescent emission \citep[e.g.][]{Burgasser2005,Driessen2022,Wandia2025} and  flaring emission   \citep[e.g.][]{Andersson2022} both attributed to gyrosynchrotron processes have been reported. Similarly, synchrotron radiation has been observed from radiation belts of later spectral type M dwarfs \citep[][]{Kao2023,Climent2023}. Coherent emission attributed to either plasma emission or ECME from the corona and or originating from large-scale magnetic fields driving auroral emission \citep[e.g.][]{Villadsen2019,Callingham2021,Gloudemans2023,Pineda2023,Smirnov2025} has also been detected. Significant insights can be drawn from these detections, e.g. the magnetic field strengths of the emitting region or object, physical source sizes, geometry of the emitting objects and electron number densities of the plasma.  Radio emission can also be generated by the interaction between the stellar wind and a hosted planet's magnetosphere. This emission arises when the ionised component of the incident stellar wind from the host star impinges on the magnetosphere and magnetic reconnection occurs at the magnetopause. Particles are accelerated along the planet's magnetic field lines and funnelled towards the planet's poles. Magnetic mirroring then reflects back particles with large pitch angles, leading to emission \citep[see][and references therein]{Callingham2024}. This mechanism is routinely seen on Earth especially on the dayside of the magnetosphere.  In situ measurements of radio power from the magnetospheres of five Solar system planets, Earth and the four gas giants have led to the formulation of a power scaling relation referred to as the radiometric \textcolor{black}{Bode's} law (RBL) which is useful in estimating the flux densities of extra-solar systems \citep[][]{Farell1999,Zarka2001,Lazio2004,Gorman2018}.

\textcolor{black}{Estimating} the magnetic field strengths and understanding the magnetic environment of M dwarfs  is both critical for habitability \textcolor{black}{ e.g. planets orbiting highly active M dwarfs may experience significant atmospheric erosion before achieving stability, potentially rendering them uninhabitable \citep[see][]{Venot2016,Nicholls2023}} and in disentangling stellar contamination from the true atmospheric signals. 
Targeted spectroscopic studies of planets orbiting nearby M dwarfs have revealed a new class of temperate ocean worlds called "Hycean Worlds" \citep[][]{Madhu2021}. In particular, potential biosignatures, carbon dioxide and methane \citep{Madhu2023}, have been detected in the atmosphere of one of these exoplanets, K2-18 b, using the James Webb Space Telescope.  A tentative detection of the abiotic biosignatures dimethyl sulfide (DMS) and dimethyl disulfide (DMDS) has also been recently claimed by \cite{Madhusudhan2025}. The claimed detection of biosignatures from K2-18 b makes the host star an interesting candidate to probe for radio emission. In this manuscript, we describe the K2-18 system in section \ref{sec:system}. In section \ref{sec:radio_data}, we discuss the VLA observations and data analysis. In section \ref{sec:radio_emission}, we discuss some mechanisms through which radio emission is generated. Finally, we present our conclusions in section \ref{sec:conclusion}.

\section{The K2-18 System}
\label{sec:system}

The K2-18 system consists of a $2.4\pm0.6$ Gyr old \citep[][]{Guinan2019} M dwarf star  of spectral type M2.5 \cite{Schweitzer2019} at a distance $\sim38.1$ pc from the Earth \citep[][]{Lindegren2021}  and hosts two exoplanets, K2-18 b \citep[][]{Montet2015} and K2-18 c \citep[][]{Cloutier2017}. K2-18 b is a sub-Neptune orbiting at a distance of $\sim0.14$ AU from the star with an orbital period  $\sim33$ days \citep[][]{Sarkis2018} and is hypothesised to be either tidally locked or in a state of spin-orbit resonance \citep[][]{Charnay2021}. The physical parameters of the star and planet are listed in Table~\ref{table:physical_params}. 
\textcolor{black}{Recent analysis of XMM-Newton observations place the soft X-ray (0.2-2.0 keV) flux at $F_\text{X}\sim 5.89\times10^{-15} \ \text{erg cm}^{-2}\ \text{s}^{-1}$ corresponding to an X-ray luminosity of $L_\text{X}\sim27.0\ \text{erg}\ \text{s}^{-1}$ consistent with a very faint emitter. This inactivity is consistent with the findings of }\cite{Modi2023} who reveal that the planet has retained its atmosphere despite erosion from extreme ultraviolet (EUV) irradiation. These conclusions are strengthened by earlier investigations on the escape rate of the planet's atmosphere by \citet[][]{dosSantos2020} who demonstrated that the planet will lose $<1\%$ of its mass over its lifetime. \textcolor{black}{We note that the low quiescent X-ray flux of the system does not preclude the occurrence of sporadic large flares and the presence of magnetic phenomena detectable at radio wavelengths. }



\begin{table}
    \renewcommand{\arraystretch}{1.4}
    \begin{tabularx}{\columnwidth}{cccc} 
		\hline
		 Property &  K2-18 & K2-18 b & Reference \\
         \hline
         SpT & M2.5 & -- & $1$ \\
         Distance (pc) & 38.1 & 38.1 & $2$ \\
         Mass ($M_\odot$, $M_J$) & 0.4951±0.0043 & $\sim0.02807$ & $3$ \\
         Radius ($R_\odot$, $R_\text{J}$) & 0.4445±0.0148 & $0.211\pm0.20$	 & $3$ \\  
         $T$ ($T_{\rm{eff}}$, $T_\text{eq}$) (K) & 3457±39 & 254.9±3.9 & $4$ \\
         $P_{\rm{rot}}$ (days) & $39.63\pm0.50$ & -- & $4$ \\
         Orbital period (days) & -- & $\sim32.939623$ & $4$ \\
         Orbital distance (AU) & -- & $\sim0.1429$ & $4$ \\
         Age (Gyr) & $2.4\pm0.6$  & -- & $5$ \\ 
         $F_\text{X} \ (\ \text{erg cm}^{-2}\ \text{s}^{-1})$  & $5.89\times10^{-15}$ & -- & $6$ \\
         \hline
	\end{tabularx}
    \caption{Physical parameters of the K2-18 system. References: $^1$\protect\cite{Schweitzer2019}, $^2$\protect\cite{Lindegren2021}, $^3$\protect\cite{Sarkis2018}, $^4$ \protect\cite{Benneke2019}, $^5$\protect\cite{Guinan2019}
    $^6$\protect\cite{Rukdee2025}}
    \label{table:physical_params}

\end{table}

\section{Radio Data}
\label{sec:radio_data}

\subsection{Observations}

We observed the K2-18 system using the VLA which consists of 27 antennas arranged into three arms, each with nine antennas, to form a Y shape. The observations were conducted over $\sim12$ weeks from 29/09/2023 to 21/12/2023 sparsely sampling the planet's orbit as a part of a technosignature search (Tremblay, C.D. et.al in prep)
utilising the Commensal Open-Source Multimode Interferometer Cluster \citep[COSMIC; ][]{COSMIC2024} and the standard \texttt{WIDAR} back end.

The bulk of the observations were conducted at S-band (2--4\,GHz) over two sessions separated by nine days. The first observation used the VLA in a non-standard hybrid configuration as the array was being reconfigured from the A to D (A$\rightarrow$D) configuration resulting in seven antennas in the A-configuration and the rest in the D-configuration. The second observation was conducted using the standard D-configuration. In the hybrid configuration, the resultant A configuration maximum baseline length in wavelength $\lambda$ at S-band is  $406\ \text{k}\lambda$ , a factor of $\sim$ 40 larger than the corresponding longest baseline of the D-configuration . The baseline mismatch leads to a pronounced change in the u-v coverage that is not reflected in the integration time resulting in a significantly distorted point spread function (PSF). 

The observation setup included a polarisation leakage calibrator J1407+2827 which was observed at the start of each observation. The phase, bandpass, and amplitude calibrator J1120+1420 was interleaved with observations of the target K2-18. Finally, 3C286 a standard VLA flux and polarisation angle calibrator was observed.  For each session, the target was observed over six scans each $\sim14$ minutes long yielding an on target time of $\sim84$ minutes. The remaining observations were short ten-minute (on-source) single scan snapshots using the standard A and D configurations in the S, C (5.5--7.5\,GHz), and X (8--10\,GHz) bands. Four snapshots were obtained in each of the frequency bands when the array was in the D-configuration. When the array had changed to the A configuration, two snapshots were obtained at S-band and a single snapshot at the C and X-bands. The calibrator setup was similar to the main dataset except the polarisation leakage calibrator was not observed. All the data were digitised at 8-bits with the correlator in continuum mode. A summary of the observations is listed in Table~\ref{table:obs_summary}.

We highlight that S-band observations using the D-configuration of the VLA are affected by source confusion due to the large synthesised beams. This results in blending of multiple faint sources that cannot be resolved individually. The array is then effectively confusion limited with the lower noise bound set by the confusion noise e.g. the confusion limit is $\sim12\ \mu$Jy $\text{beam}^{-1}$ \footnote{\href{https://obs.vla.nrao.edu/ect/}{VLA Exposure Calculator}} for S-band D-configuration observations. Confusion in the other configurations and at higher frequencies is negligible. In addition to confusion noise, S-band  D-configuration observations   are significantly impacted by radio frequency interference (RFI).

\begin{table*}
  \centering
    \renewcommand{\arraystretch}{1.1}
    \begin{tabular}{cccccccc} \\
    \hline
      \noalign{\vskip 3pt}  

     Obs.Date & Band & Configuration & \makecell{Time on Source \\ (min)} & \makecell{Restoring Beam \\ (arcsec$\times$arcsec)} & \makecell{Self-calibration \\ Parameters} & \multicolumn{2}{c}{\makecell{$1\sigma$ rms  ($\mu\text{Jy\,beam}^{-1}$)}}  \\  
     & & & & & & Stokes I & Stokes V \\

    \noalign{\vskip 3pt}   %
    
    \hline
    \noalign{\vskip 2pt}   %
    29/09/2023 &  C &  A & 10 & 1.0$\times$0.3 &  \makecell{5 p only (inf, 148s, 36s,8s, int) and 1 ap (inf)} & 14.0 & 12.2 \\
    29/09/2023 & X &  A & 10 & 0.5$\times$0.2 & \makecell{5 p only (inf, 148s, 36s,8s, int)} & 11.5 & 9.8\\
    30/09/2023 & S &  A & 10 & 1.3$\times$0.6 & \makecell{5 p only (inf, 148s, 36s,8s,int)} & 16.6  & 13.3 \\   
    30/09/2023 & S &  A & 10 & 1.0$\times$0.6  & \makecell{4 p only (inf, 148s, 36s,8s)} & 15.0 & 11.5 \\   
    
    13/10/2023 & S & A$\rightarrow$D & 84 & 6.5$\times$1.3 & ---  & 71.3 & 16.0 $^a$ \\
    22/10/2023 & S &  D & 84 &  $19.2\times17.3$  & \makecell{2 p only (inf, 166s) and 1 ap (inf)} & 29.0 & 9.7 \\
    01/12/2023 & S &  D & 10 & 22.4$\times$19.1 &  \makecell{4 p only (inf, 148s, 36s,8s) and 1 ap (inf)} & 30.1 & 22.5 \\
    
    01/12/2023 & C &  D & 10 & 18.2$\times$8.9 & \makecell{3 p only (inf, 148s, 36s) and 1 ap (inf) } & 16.5 & 12.2 \\
    01/12/2023 & X &  D & 10 & 16.8$\times$6.2 & \makecell{1p only (inf)} & 44.8 & 35.5 \\
    08/12/2023 & S &  D & 10 & 36.0$\times$19.0 & \makecell{5 p only (inf, 148s, 36s,8s, int) and 1 ap (inf)} & 51.9 & 21.9 \\
    08/12/2023 & C &  D & 10 & 15.2$\times$9.2  & \makecell{3 p only (inf, 148s, 36s) and 1 ap (inf)} & 14.1 & 10.3 \\
    08/12/2023 & X &  D & 10 & 10.8$\times$6.8 & \makecell{1p only (inf)} & 13.6 & 9.0 \\
    13/12/2023 & X &  D & 10 & 11.3$\times6.6$ & \makecell{3 p only (inf, 148s, 36s) and 1 ap (inf) } & 13.4 & 10.6 \\ 
    14/12/2023 & S &  D & 10 & 25.1$\times$18.5  & \makecell{5 p only (inf, 148s, 36s,8s, int) and 1 ap (inf)} & 46.5 & 14.9 \\
    14/12/2023 & C &  D & 10 & 12.6$\times$9.4& \makecell{5 p only (inf, 148s, 36s,8s, int) and 1 ap (inf)} & 14.3 & 10.2\\
    18/12/2023 & X &  D & 10& 12.3 $\times$6.7 & \makecell{3 p only (inf, 148s, 36s) and 1 ap (inf) } & 11.2 & 9.3\\
    19/12/2023 & C &  D & 10 & 13.3$\times$9.3 & \makecell{2 p only (inf, 148s) and 1 ap (inf)} & 13.2 & 9.5 \\
    21/12/2023 &  S &  D & 10 & 36.5$\times$16.9  & \makecell{3 p only (inf, 148s, 36s) } & 70.3 & 27.3  \\
    \noalign{\vskip 3pt}
    
    \hline
	\end{tabular}
    \caption{List of observation of K2-18 ordered by date. The bands are defined by their designated frequency ranges, the S-band covers 2-4 GHz, C-band 5.5-7.5 GHz and X-band 8-10 GHz. The observations were conducted using the A and D configuration of the VLA. The bulk of the data were observed at S-band over an observing period of 84+84 minutes. The rest of the observations were short 10 minute scans. The restoring beam for each observation obtained from self calibrated images is also presented. The details of the self-calibration parameters are presented with p denoting phase-only self-calibration and ap denoting amplitude-and-phase self-calibration. The solution intervals are also reported where inf corresponds to the full scan length and int conrresponds to the integration time. The integer preceding p or ap specifies the self-calibration rounds with the last iteration indicating convergence. 
    $^a$ In an attempt to shape the PSF, a uniformly weighted (robust=-1) Gaussian taper of beam width equivalent to the B-configuration VLA resolution at S band ( $\sim$2.1\arcsec) was applied to the visibilities, nonetheless, self-calibration was unsuccessful. }
    \label{table:obs_summary}
\end{table*}

\subsection{Calibration}

We processed the data using the VLA Calibration Pipeline 2024.1.0.8 \footnote{\href{https://science.nrao.edu/facilities/vla/data-processing/pipeline}{VLA Calibration Pipeline 2024.1.0.8}} which is integrated into the Common Astronomy Software Applications \citep[CASA; ][version 6.6.1]{McMullin2007,Bean2022}. The pipeline first imports the data stored in a Science Data Model-Binary Data Format [SDM-BDF]) to a \textcolor{black}{MeasurementSet (MS)}, performs Hanning smoothing to reduce Gibbs ringing, applies online flags, and performs various calibration steps ranging from antenna position corrections to gain and phase calibrations. Automatic flagging of radio frequency interference (RFI) is also executed. Finally, the data are science ready. We note that plots are made at various steps to serve as a diagnostic for the calibration quality. For the two 84 min S-band observations, J1407+2827 was used to correct for polarisation leakage between the feeds and 3C286 was used to calibrate the absolute position of the polarisation angle. Polarisation calibration for the snapshot observations was not possible as a polarisation leakage calibrator was not available.

\subsection{Imaging and Self-calibration}


To assess the quality of the data calibration, we produced images using the CASA task \texttt{tclean}. We employed the multi-term multi-frequency \citep[MTMFS; ][]{Rau2011} deconvolver to account for the large observing bandwidth (2\,GHz). The images were also  Briggs weighted \citep[][]{Briggs1995} with a robust parameter of 0.5 (biasing slightly toward natural weighting). We observed no distinct emission at the position of the K2-18 system. In the images, the noise significantly deviated from the expected  noise floor set by the confusion limit for S-band D-configuration observations and the thermal noise for the rest of the observations. Upon closer examination, the cause of the elevated noise was determined to be QSO B1127+078 a bright quasar located $\sim3$ \arcmin from the target position \citep[][]{Petrov2024} at a peak flux density in the D-configuration of $\sim50$ mJy $\text{beam}^{-1}$, $\sim22$ mJy $\text{beam}^{-1}$ and $\sim10$ mJy $\text{beam}^{-1}$ at S, C and X-bands, respectively. Imaging artifacts associated with the quasar significantly limited the dynamic range of the images.

To improve the fidelity of the images,  we performed self-calibration using the official pipeline self-calibration scripts \footnote{\href{https://github.com/jjtobin/auto_selfcal}{https://github.com/jjtobin/auto\_selfcal}} which we edited to fit our needs. For each band, we produced a wide-field image of size $1.5$ times the primary beam at S-band and covering the entire primary beam at C and X-bands. All the images encompassed the target, the bright quasar and multiple faint sources in the field of view. We then performed self-calibration for each of the observations using the parameters listed in Table~\ref{table:obs_summary}. We note the S-band D-configuration data are significantly degraded by RFI and require extensive flagging resulting in noise levels significantly above the confusion limit of $\sim12\ \mu \text{Jybeam}^{-1}$. The noise levels for the 10 minute snapshots at C and X-bands are consistent with the theoretical thermal noise level of $\sim9\ \mu \text{Jybeam}^{-1}$ for a dual polarisation Stokes I Briggs weighted image.  The noise levels for all the observations in the Stokes I and V are listed in Table~\ref{table:obs_summary}. Attempts at self-calibration were unsuccessful for the data observed using the hybrid configuration due to the poorly characterised PSF. Excluding the seven antennas in the A configuration did not mitigate the issue, as the data were extensively flagged leaving only a small ($<5$\%) unusable subset. Similarly, tapering the $uv$-coverage and changing the weighting scheme did not alleviate the problem. Consequently, the data were excluded from further analysis.

\subsection{Variability Analysis}

To search for short duration bursts, we began by subtracting the CLEAN components stored in the \texttt{modelcolumn} of the \textcolor{black}{MS} from the self calibrated visibilities using the CASA task \texttt{uvsub}. Since the target was not detected, it was not included in the self-calibration mask and thus not deconvolved. Consequently, the \texttt{modelcolumn} contained no associated CLEAN components precluding any risk of subtraction. To subtract faint $\mu$Jy sources that were otherwise not included in the self-calibration model, we generated large images similar in size to those discussed in the preceding section, masked the target position, deconvolved the images in the Stokes I and V using \texttt{WSClean} \citep[][]{WSClean2014} and similarly subtracted the CLEAN components from the visibilities using \texttt{uvsub}.
We then used the \texttt{table} and \texttt{ms} tools available via the \texttt{casatools} package to parse the \textcolor{black}{MS} after which we averaged all baselines, frequency channels and spectral windows and binned the data at a cadence of one minute yielding a single visibility per bin. Subsequently, we produced light curves for the C and X-band observations. Light curves for the S-band observations were not produced due to extensive flagging of the data in addition to difficulties in accurately subtracting the numerous faint background sources. 
\\~\\

\section{Radio emission}

\label{sec:radio_emission}

\begin{figure*}
  \centering
  \setlength{\tabcolsep}{-18pt} 
  \renewcommand{\arraystretch}{0} 
  \begin{tabular}{cccc}
  
    \includegraphics[width=0.38\textwidth]{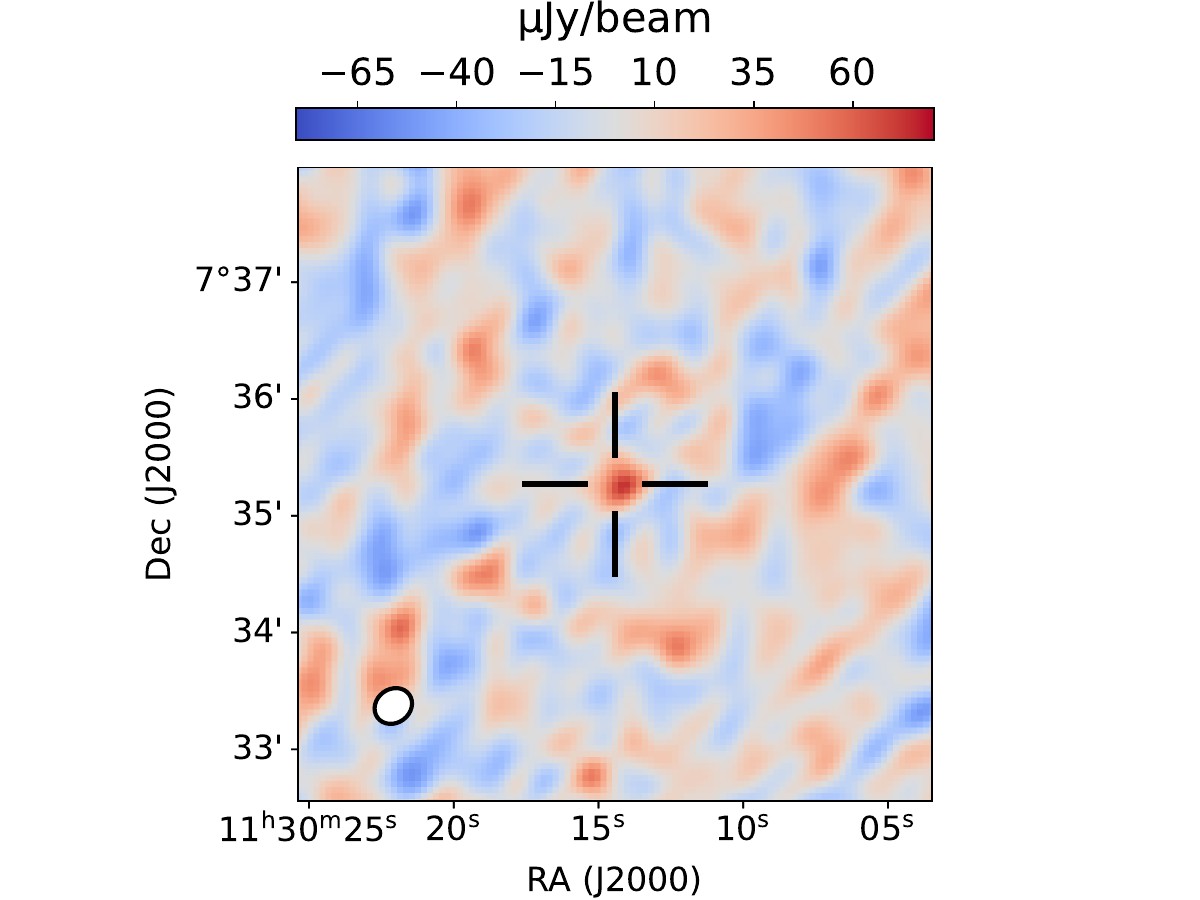} &
    
    \includegraphics[width=0.38\textwidth]{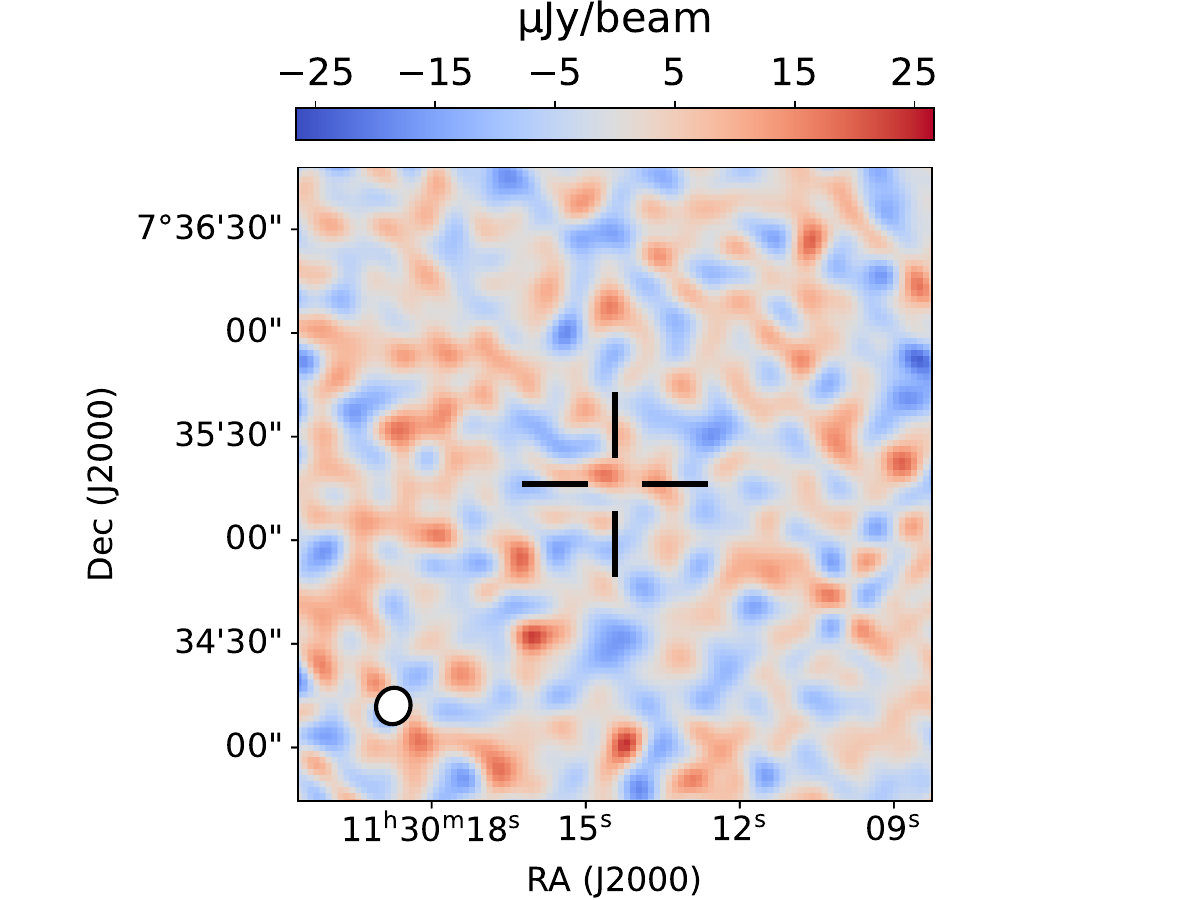} \vspace{0.25cm}  & 
    
    \includegraphics[width=0.38\textwidth]{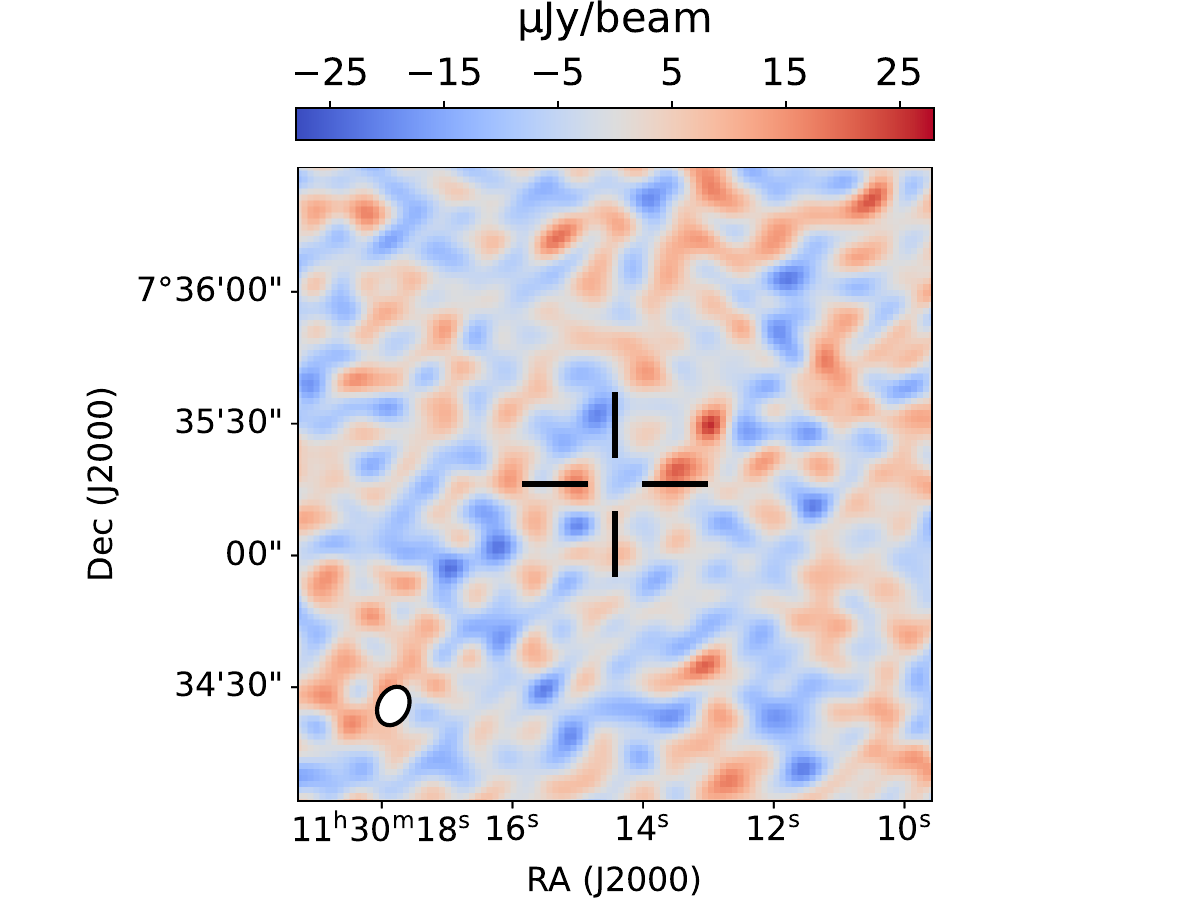} \vspace{0.1cm}
    \\
    \vspace{0.3cm}
    \includegraphics[width=0.38\textwidth]{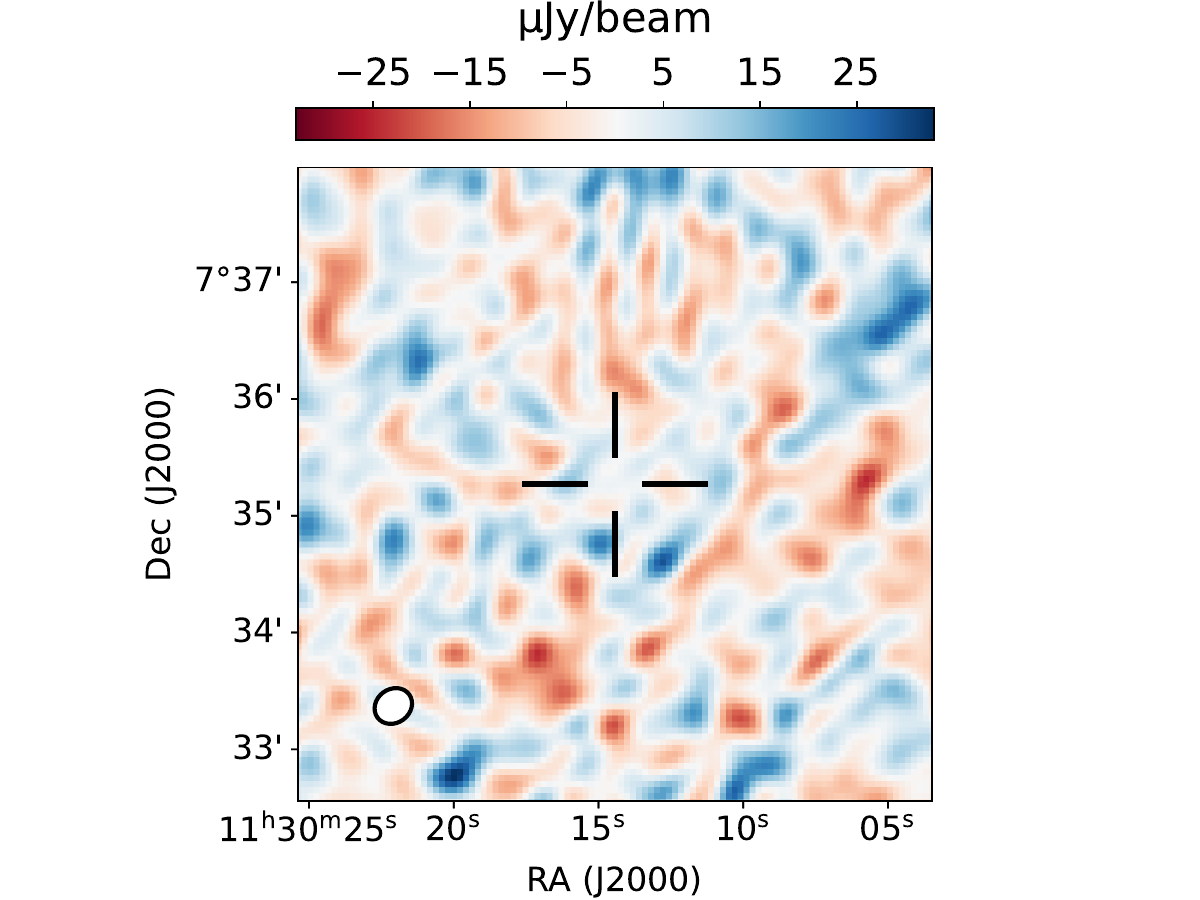} &
      \includegraphics[width=0.38\textwidth] {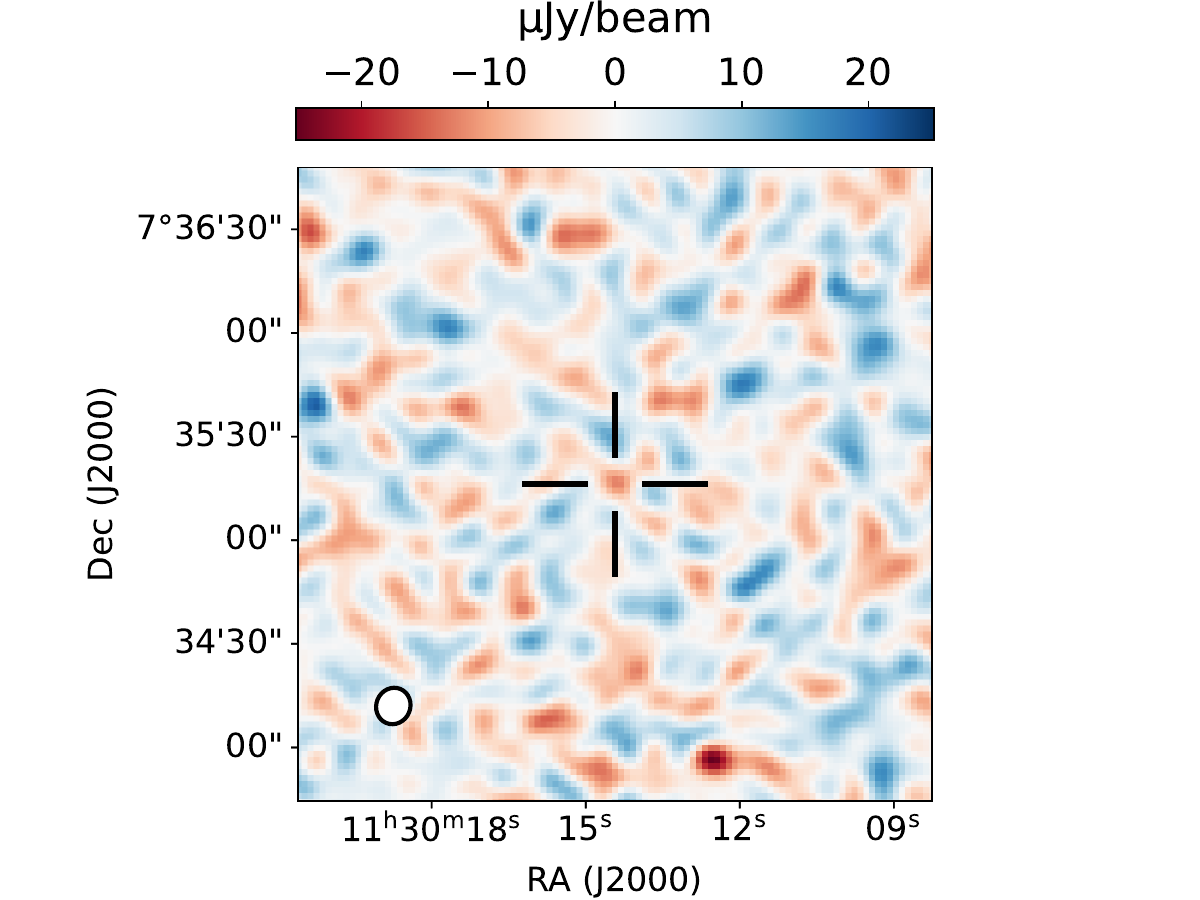}  &
    \includegraphics[width=0.38\textwidth]  
   {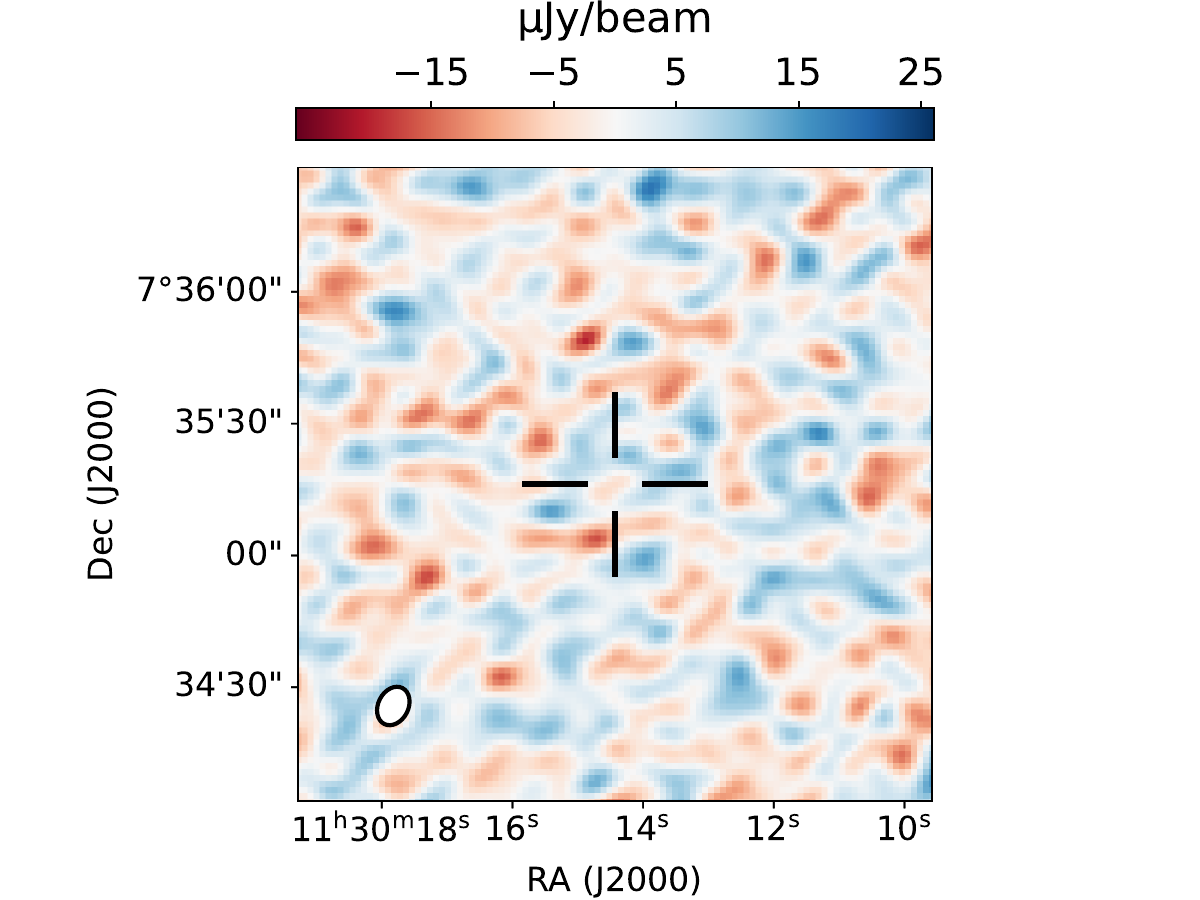} 
     \vspace{0.1cm} \\ (a) & (b) & (c) \vspace{0.1cm}
    

    \end{tabular}
    \caption{ \textcolor{black}{Panel (a): S-band Stokes I image (top) and Stokes V (bottom). The  Stokes I $1\sigma$ rms noise is $16.6\  \mu \text{Jybeam}^{-1}$ and the corresponding $1\sigma$ rms noise of the Stokes V image is $7.1\  \mu \text{Jybeam}^{-1}$. Panel(b) C-band Stokes I image (top) and Stokes V (bottom). The Stokes I $1\sigma$ rms noise is $5.9\  \mu \text{Jybeam}^{-1}$ and the corresponding $1\sigma$ rms noise of the Stokes V image is $5.0\  \mu \text{Jybeam}^{-1}$. Panel(c) X-band Stokes I image (top) and Stokes V (bottom). The Stokes I $1\sigma$ rms noise is $6.0\  \mu \text{Jybeam}^{-1}$ and the corresponding $1\sigma$ rms noise of the Stokes V image is $5.2\  \mu \text{Jybeam}^{-1}$. 
    The \textcolor{black}{black} cross in the images indicates the position of K2-18. The restoring beam is shown in the filled white circle on the bottom left of each image. All the images are produced from observations conducted using the D-configuration of the VLA.}}

  \label{fig:maps}
\end{figure*}

\subsection{\textcolor{black}{Planetary Auroral Emission}}
\label{section:stellar_wind_magnetosphere}

The RBL described in section~\ref{sec:intro} is use to determine the median power output caused by the impinging ionised stellar wind on the planet's magnetosphere. The emission is at a central frequency $\nu_c$ presented by \cite{Farell1999} as

\begin{equation}
    \nu_c \sim 23.5\ \left(\frac{\omega_\text{P}}{\omega_\text{J}}\right)\ \left(\frac{M_\text{P}}{M_\text{J}}\right)^{5/3} \left(\frac{R_p}{R_\text{J}}\right)^3\ \ \ \text{MHz}\ ,
    \label{eqn:central_freq}
\end{equation}

and the median emitted radio power $P_\text{rad}$ as \citep[][]{Farell1999}

\begin{equation}
\begin{gathered}
    P_{\rm{rad}} = 4\times10^{18} \  
    \left(\frac{\omega_\text{P}}{\omega_\text{J}}\right)^{0.79}\left(\frac{a_\text{p}}{a_J}\right)^{-1.60}\left(\frac{M_\text{P}}{M_\text{J}}\right)^{1.33}\ \rm{erg}\ \rm{s}^{-1}\  \ .
    \label{eqn:rbl}
\end{gathered}
\end{equation}

In equations~\ref{eqn:central_freq} and~\ref{eqn:rbl}, $\omega$, $a$, $M$ and $R$ are the rotation period, orbital distance, mass, and radius. The subscripts ``P" and ``J" indicate values for the planet and Jupiter, respectively. Using equation~\ref{eqn:central_freq}, we estimate $v_\text{c}\sim$45 kHz which is below the plasma frequency $\left(\sim10\ \text{MHz}\right)$ of the Earth's ionosphere, and therefore not detectable from ground based radio observatories. We determine a median radio power output of  $P_\text{rad}\sim3.2\times10^{20}\ \text{erg s}^{-1}$, and a median flux, $F_\nu\sim8.2\times10^{-26}\ \text{erg\ s}^{-1}\ \text{cm}^{-2}\ \text{Hz}^{-1}$ or $\sim$8.2 mJy, using $F_\nu=P_\text{rad}/4\pi d^2 \Delta \nu$, where $\Delta\nu=0.5\nu_c$ \citep[see][]{Farell1999}. We note that we have assumed isotropic beaming in the evaluation of the flux density. Although such fluxes will be accessible to future space observatories, the frequencies of the emission may remain inaccessible due to free-free absorption of the radio waves along the line of sight by the interstellar medium, especially for distant star-exoplanet systems \citep[see][]{Burkhart2017}.

\subsection{\textcolor{black}{Coronal Gyrosynchrotron Emission}}

Radio emission from the hot corona of K2-18 is likely to occur via gyrosynchrotron processes \textcolor{black}{as is typical for M dwarfs}. To search for slowly varying emission arising from gyrosynchrotron processes, we combined data for each observing epoch and band and generated images and made no detection. Using the more sensitive (due to more snapshots) D-configuration observations, we placed $3\sigma$ Stokes I and V constrains of $49.8\  \mu \text{Jybeam}^{-1}$ and  $21.3\  \mu \text{Jybeam}^{-1}$ at S-band , $17.7\  \mu \text{Jybeam}^{-1}$ and  $15.0\  \mu \text{Jybeam}^{-1}$ at C-band  and $18.0\  \mu \text{Jybeam}^{-1}$ and  $15.6\  \mu \text{Jybeam}^{-1}$ at X-band. We present the combined images showing the $1\sigma$ rms for each observing band in Figure~\ref{fig:maps}.

To place limits on the luminosities, we first estimate the bolometric luminosity $L_\text{bol}$ using the Stefan Boltzmann's law, $L_\text{bol}=4\pi \sigma_\text{sb} R^2T_\text{eff}^4$, where $\sigma_\text{sb}$ is the Stefan Boltzmann's constant and $R$ and $T_\text{eff}$ are the radius and effective temperature of K2-18. We determine $\log{L_\text{bol}}\sim32.0 \ \text{erg}\ \text{s}^{-1}$ using the parameters in Table~\ref{table:physical_params}. For the spectral luminosity $L_\nu = 4\pi d^2S_\nu$ where $d$ is the distance to K2-18 and $S_\nu$ is the flux density, we estimate  $\log{L_\nu}<$ 14.0 $\text{erg s}^{-1}\text{Hz}^{-1}$ at S-band, <13.5 $\text{erg s}^{-1}\text{Hz}^{-1}$  at C-band and <13.5 $\text{erg s}^{-1}\text{Hz}^{-1}$ at X-band. The corresponding radio luminosities $L_\text{R} = \int L_\nu \rm{d}\nu$, where $\rm{d}\nu$ is the observing bandwidth are $\log{L_\text{R}}<$ 23.2 $\text{erg s}^{-1}$ at S-band, <22.8 $\text{erg s}^{-1}$ at C-band and <22.8 $\text{erg s}^{-1}$ at X-band and determine a ratio of the radio to bolometric luminosities of $\log L_\text{R}/\log L_\text{bol}<-8.8$ for the three observing bands.

\textcolor{black}{Using the X-ray luminosity of $\log{L_\text{X}}\sim27.0\ \text{erg s}^{-1}$ and the measured C band (5 GHz) upper limits on the radio luminosity, we establish the luminosity is possibly in agreement with the Gudel-Benz relationship which relates the soft X-ray emission to the 5 GHz radio luminosity as follows; $\text{log} L_\text{X} \lesssim\text{log}L_\text{R}+15.5$ \citep[][]{GudelBenz1993}}.

\subsection{\textcolor{black}{Electron Cyclotron Maser Emission}}
\begin{figure}
    \centering
    \includegraphics[width=\linewidth]{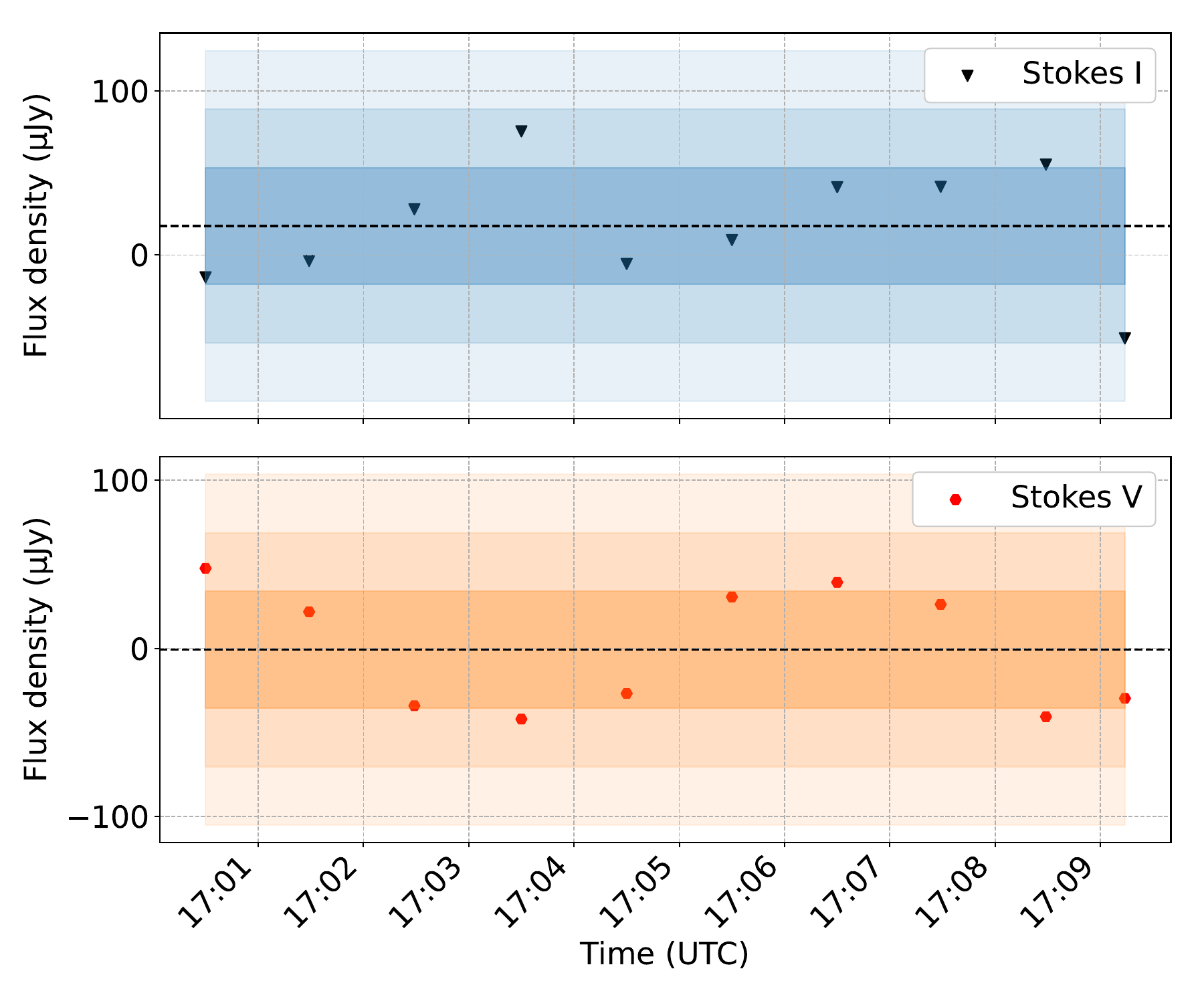}
    \caption{C-band light \textcolor{black}{curves} of K2-18 for the observation conducted on 08/12/2023 using the VLA in its D-configuration . The visibilities are binned at a cadence of 1 minute. The broken line represents the mean values in the Stokes I and V at $\sim17.6\ \mu \text{Jy}$ and $\sim-0.8\ \mu \text{Jy}$, respectively. The rms noise $\sigma$ is $\sim35.4\ \mu \text{Jy}$ and $\sim34.8\ \mu \text{Jy}$ in the Stokes I and V and is represented by the shaded region with decreasing intensity corresponding to 1, 2 and $3\sigma$ respectively.}
    \label{fig:c_band_light_curve}
\end{figure}

\begin{figure}
    \centering
    \includegraphics[width=\linewidth]{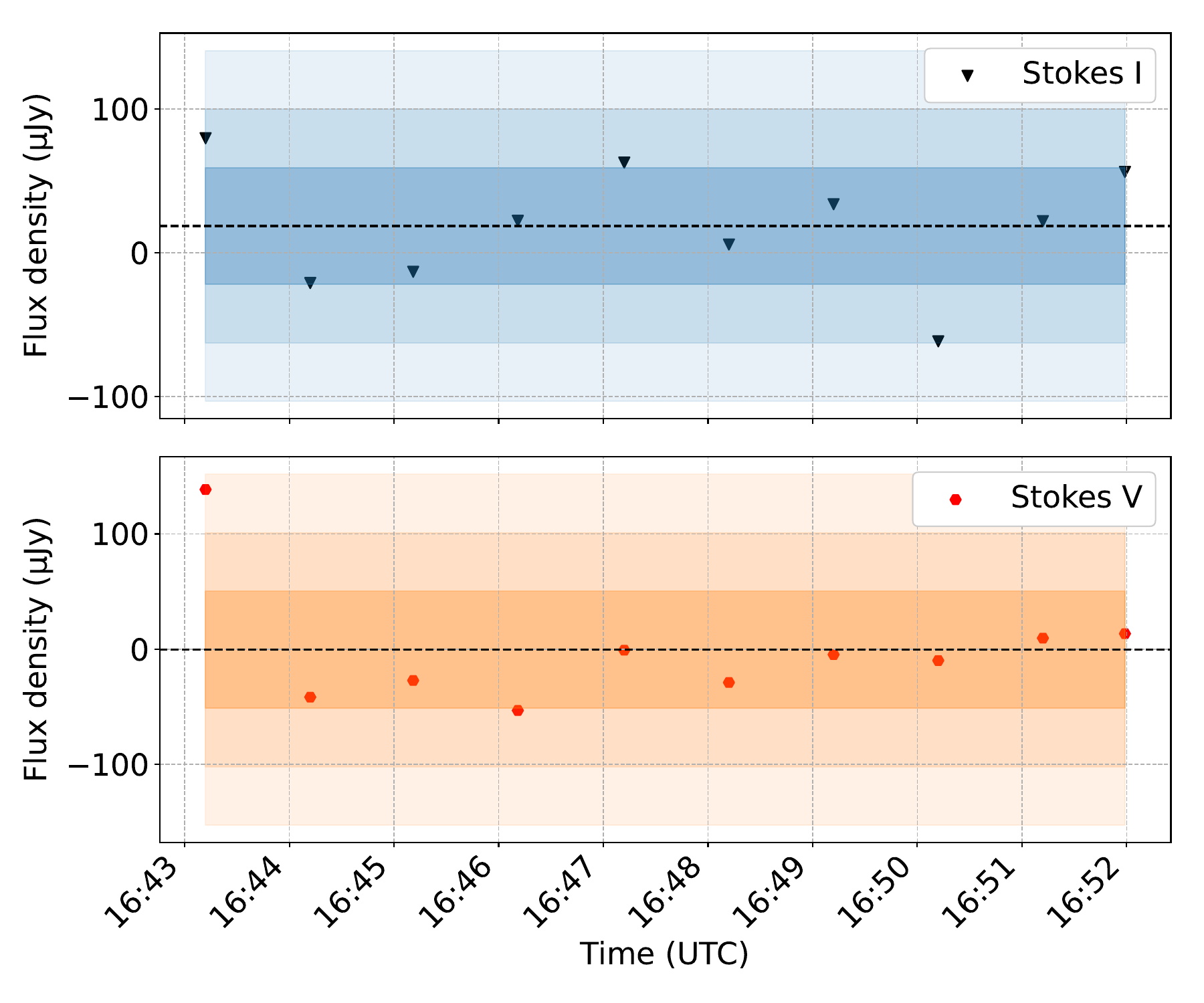}
    \caption{X-band light \textcolor{black}{curves} of K2-18 for the observation conducted on 18/12/2023 using the VLA in its D-configuration . The visibilities are binned at a cadence of 1 minute. The broken line represents the mean values in the Stokes I and V at $\sim18.7\ \mu \text{Jy}$ and $\sim-0.6\ \mu \text{Jy}$, respectively. The rms noise $\sigma$ is $\sim40.6\ \mu \text{Jy}$ and $\sim50.8\ \mu \text{Jy}$ in the Stokes I and V and is represented by the shaded region with decreasing intensity corresponding to 1, 2 and $3\sigma$ respectively.
    }
    \label{fig:x_band_light_curve}
\end{figure}

We searched for short duration bursts which are characteristic of ECME due to coronal or auroral emission by generating light curves from the C and X-band observations. We present the light curves in Figures~\ref{fig:c_band_light_curve} and~\ref{fig:x_band_light_curve} . We did not detect any emission at the location of K2-18 above $3\sigma$, where $\sigma$ is the rms noise in an observing cadence of one minute . Given the null detection of K2-18, we present one example per band by selecting the dataset with the best self-calibration and lowest noise in the Stokes I and V to demonstrate the absence of a detectable signal.


\section{Conclusions}
\label{sec:conclusion}
The James Webb Space Telescope, through transmission spectroscopy, has opened a new window into the study of exoplanetary atmospheres.  Over its mission lifetime, the telescope is expected to characterise the atmospheres of some of the nearby stars, many of which are M dwarfs. The detection of biosignatures is within sight, and a unified approach involving radio observations is necessary to truly constrain habitability. Radio observations will also prove crucial to constraining the magnetic properties of M dwarfs, by providing direct measurements of magnetic field strengths and potentially activity levels by detecting radio emission associated with magnetic phenomena. 

In this manuscript,we have analysed VLA radio observations conducted over 12 weeks at a frequency bandwidth ranging from $2-10$ GHz and reported null detections for slowly varying emission associated with gyrosynchrotron processes. We have placed $3\sigma$ upper limits of $49.8\ \mu\text{Jybeam}^{-1}$ at S-band, $17.7\ \mu\text{Jybeam}^{-1}$ at C-band and $18\ \mu\text{Jybeam}^{-1}$  at X-band. We similarly place $3\sigma$ upper limits for the spectral luminosity at $\log{L_\nu}<14.0\ \text{erg s}^{-1}\text{Hz}^{-1}$ and a ratio of the radio to bolometric luminosity of $\log{L_\text{R}}/\log{L_\text{bol}}<-8.8$ where $\log{L_\text{bol}}$ is the total bolometric luminosity at $\sim32\ \text{erg\ s}^{-1}$. We have further searched for short duration bursts from ECME using light curves produced from C and X-band observations and made no detection at a $3\sigma$ level. \textcolor{black}{These constraints are vital for building realistic models of stellar surface inhomogeneities and mitigating their impact on the interpretation of transmission spectra. In particular, the null detection of incoherent emission associated with gyrosynchrotron emission may imply a low flaring rate for K2-18, translating to lower temporal variability in the transmission spectra and a low filling rate for spots and faculae. These findings are supported by recent analysis of X-ray data from the system by \cite{Rukdee2025} and may suggest that planetary transmission spectra from K2-18 b are free from contamination, nevertheless, we advise caution in their interpretation owing to the sparse sampling of the planet's orbit in our observations.}

We highlight that there are several factors affecting detectability of radio emission from these mechanisms. The viewing angle of the instrument is particularly important for ECME driven emission  which requires favourable observing angles. The magnetic field strength also affects the emission frequency for both ECME gyrosynchrotron and physical properties of the plasma e.g high number densities lead to suppression of gyrosunchrotron radio emission due to Razin-Tsytovich effect \citep[][]{Dulk1985} and similarly very dense plasma inhibit the propagation of radio waves associated with ECME \citep[see][]{Gudel2002}. We also establish that radio emission from the planet's magnetosphere is undetectable with current Earth-bound instruments.

\section{Acknowledgements}
This project has been made possible in part by a grant from the SETI Institute. This work made
use of Astropy:3 a community-developed core Python package and an
ecosystem of tools and resources for astronomy \citep[][]{astropy2013,astropy2018,astropy2022}. This research has made use of the NASA
Exoplanet Archive, which is operated by the California Institute of
Technology, under contract with the National Aeronautics and Space
Administration under the Exoplanet Exploration Program. The National Radio Astronomy Observatory is a facility of the National Science Foundation
operated under cooperative agreement by Associated Universities,
Inc. We thank the anonymous referee for their useful comments which have substantially improved the manuscript.

\section*{Data Availability}

Data underlying this article are publicly available in the NRAO Data Archive at \url{https://data.nrao.edu/portal} and can be accessed with project code 23B-307.


\bibliographystyle{mnras}
\bibliography{example} 

@ARTICLE{Wandia2025,
       author = {{Wandia}, Kelvin and {Garrett}, Michael A. and {Beswick}, Robert J. and {Radcliffe}, Jack F. and {Gajjar}, Vishal and {Williams-Baldwin}, David and {Tremblay}, Chenoa and {McDonald}, Iain and {Andersson}, Alex and {Siemion}, Andrew},
        title = "{Radio Emission from a Nearby M dwarf Binary}",
      journal = {arXiv e-prints},
         year = 2025,
        month = jul,
          eid = {arXiv:2507.20681},
        pages = {arXiv:2507.20681},
          doi = {10.48550/arXiv.2507.20681},
archivePrefix = {arXiv},
       eprint = {2507.20681},
 primaryClass = {astro-ph.SR},
       adsurl = {https://ui.adsabs.harvard.edu/abs/2025arXiv250720681W}
}

@INPROCEEDINGS{Briggs1995,
       author = {{Briggs}, D.~S.},
        title = "{High Fidelity Interferometric Imaging: Robust Weighting and NNLS Deconvolution}",
    booktitle = {American Astronomical Society Meeting Abstracts},
         year = 1995,
       series = {American Astronomical Society Meeting Abstracts},
       volume = {187},
        month = dec,
          eid = {112.02},
        pages = {112.02},
       adsurl = {https://ui.adsabs.harvard.edu/abs/1995AAS...18711202B}
}

@ARTICLE{Sarkis2018,
       author = {{Sarkis}, Paula and {Henning}, Thomas and {K{\"u}rster}, Martin and {Trifonov}, Trifon and {Zechmeister}, Mathias and {Tal-Or}, Lev and {Anglada-Escud{\'e}}, Guillem and {Hatzes}, Artie P. and {Lafarga}, Marina and {Dreizler}, Stefan and {Ribas}, Ignasi and {Caballero}, Jos{\'e} A. and {Reiners}, Ansgar and {Mallonn}, Matthias and {Morales}, Juan C. and {Kaminski}, Adrian and {Aceituno}, Jes{\'u}s and {Amado}, Pedro J. and {B{\'e}jar}, Victor J.~S. and {Hagen}, Hans-J{\"u}rgen and {Jeffers}, Sandra and {Quirrenbach}, Andreas and {Launhardt}, Ralf and {Marvin}, Christopher and {Montes}, David},
        title = "{The CARMENES Search for Exoplanets around M Dwarfs: A Low-mass Planet in the Temperate Zone of the Nearby K2-18}",
      journal = {\aj},
         year = 2018,
        month = jun,
       volume = {155},
       number = {6},
          eid = {257},
        pages = {257},
          doi = {10.3847/1538-3881/aac108},
archivePrefix = {arXiv},
       eprint = {1805.00830},
 primaryClass = {astro-ph.EP},
       adsurl = {https://ui.adsabs.harvard.edu/abs/2018AJ....155..257S}
}

@ARTICLE{GudelBenz1993,
       author = {{Guedel}, Manuel and {Benz}, Arnold O.},
        title = "{X-Ray/Microwave Relation of Different Types of Active Stars}",
      journal = {\apjl},
         year = 1993,
        month = mar,
       volume = {405},
        pages = {L63},
          doi = {10.1086/186766},
       adsurl = {https://ui.adsabs.harvard.edu/abs/1993ApJ...405L..63G}
}

@ARTICLE{Pohjolainen2020,
       author = {{Pohjolainen}, Silja and {Talebpour Sheshvan}, Nasrin},
        title = "{Cut-off features in interplanetary solar radio type IV emission}",
      journal = {Advances in Space Research},
         year = 2020,
        month = mar,
       volume = {65},
       number = {6},
        pages = {1663-1672},
          doi = {10.1016/j.asr.2019.05.034},
archivePrefix = {arXiv},
       eprint = {1906.07534},
 primaryClass = {astro-ph.SR},
       adsurl = {https://ui.adsabs.harvard.edu/abs/2020AdSpR..65.1663P}
}

@ARTICLE{Melrose1980,
       author = {{Melrose}, D.~B.},
        title = "{The Emission Mechanisms for Solar Radio Bursts}",
      journal = {\ssr},
         year = 1980,
        month = may,
       volume = {26},
       number = {1},
        pages = {3-38},
          doi = {10.1007/BF00212597},
       adsurl = {https://ui.adsabs.harvard.edu/abs/1980SSRv...26....3M}
}

@ARTICLE{Charnay2021,
       author = {{Charnay}, B. and {Blain}, D. and {B{\'e}zard}, B. and {Leconte}, J. and {Turbet}, M. and {Falco}, A.},
        title = "{Formation and dynamics of water clouds on temperate sub-Neptunes: the example of K2-18b}",
      journal = {\aap},
         year = 2021,
        month = feb,
       volume = {646},
          eid = {A171},
        pages = {A171},
          doi = {10.1051/0004-6361/202039525},
archivePrefix = {arXiv},
       eprint = {2011.11553},
 primaryClass = {astro-ph.EP},
       adsurl = {https://ui.adsabs.harvard.edu/abs/2021A&A...646A.171C}
}

@ARTICLE{Winters2019,
       author = {{Winters}, Jennifer G. and {Henry}, Todd J. and {Jao}, Wei-Chun and {Subasavage}, John P. and {Chatelain}, Joseph P. and {Slatten}, Ken and {Riedel}, Adric R. and {Silverstein}, Michele L. and {Payne}, Matthew J.},
        title = "{The Solar Neighborhood. XLV. The Stellar Multiplicity Rate of M Dwarfs Within 25 pc}",
      journal = {\aj},
         year = 2019,
        month = jun,
       volume = {157},
       number = {6},
          eid = {216},
        pages = {216},
          doi = {10.3847/1538-3881/ab05dc},
archivePrefix = {arXiv},
       eprint = {1901.06364},
 primaryClass = {astro-ph.SR},
       adsurl = {https://ui.adsabs.harvard.edu/abs/2019AJ....157..216W}
}

@ARTICLE{Kao2023,
       author = {{Kao}, Melodie M. and {Mioduszewski}, Amy J. and {Villadsen}, Jackie and {Shkolnik}, Evgenya L.},
        title = "{Resolved imaging confirms a radiation belt around an ultracool dwarf}",
      journal = {\nat},
         year = 2023,
        month = jul,
       volume = {619},
       number = {7969},
        pages = {272-275},
          doi = {10.1038/s41586-023-06138-w},
archivePrefix = {arXiv},
       eprint = {2302.12841},
 primaryClass = {astro-ph.EP},
       adsurl = {https://ui.adsabs.harvard.edu/abs/2023Natur.619..272K}
}

@ARTICLE{Dulk1979,
       author = {{Dulk}, G.~A. and {Melrose}, D.~B. and {White}, S.~M.},
        title = "{The gyrosynchrotron emission from quasi-thermal electrons and applications to solar flares}",
      journal = {\apj},
         year = 1979,
        month = dec,
       volume = {234},
        pages = {1137-1147},
          doi = {10.1086/157597},
       adsurl = {https://ui.adsabs.harvard.edu/abs/1979ApJ...234.1137D}
}

@ARTICLE{Climent2023,
       author = {{Climent}, J.~B. and {Guirado}, J.~C. and {P{\'e}rez-Torres}, M. and {Marcaide}, J.~M. and {Pe{\~n}a-Mo{\~n}ino}, L.},
        title = "{Evidence for a radiation belt around a brown dwarf}",
      journal = {Science},
         year = 2023,
        month = sep,
       volume = {381},
       number = {6662},
        pages = {1120-1124},
          doi = {10.1126/science.adg6635},
archivePrefix = {arXiv},
       eprint = {2303.06453},
 primaryClass = {astro-ph.SR},
       adsurl = {https://ui.adsabs.harvard.edu/abs/2023Sci...381.1120C}
}

@ARTICLE{Barclay2021_inhomogeneous,
       author = {{Barclay}, Thomas and {Kostov}, Veselin B. and {Col{\'o}n}, Knicole D. and {Quintana}, Elisa V. and {Schlieder}, Joshua E. and {Louie}, Dana R. and {Gilbert}, Emily A. and {Mullally}, Susan E.},
        title = "{Stellar Surface Inhomogeneities as a Potential Source of the Atmospheric Signal Detected in the K2-18b Transmission Spectrum}",
      journal = {\aj},
         year = 2021,
        month = dec,
       volume = {162},
       number = {6},
          eid = {300},
        pages = {300},
          doi = {10.3847/1538-3881/ac2824},
archivePrefix = {arXiv},
       eprint = {2109.14608},
 primaryClass = {astro-ph.EP},
       adsurl = {https://ui.adsabs.harvard.edu/abs/2021AJ....162..300B}
}

@ARTICLE{Smirnov2025,
       author = {{Smirnov}, O.~M. and {Golden}, A. and {Myburgh}, T. and {Ngcebetsha}, B. and {Tasse}, C. and {Heywood}, I. and {Ramaila}, A.~J.~T. and {Thompson}, M.~A. and {Kenyon}, J.~S. and {Perkins}, S.~J. and {Dawson}, J. and {Bester}, H.~L. and {Bright}, J.~S. and {Oozeer}, N. and {Samboco}, V.~G.~G. and {Sihlangu}, I. and {Choza}, C.},
        title = "{Mining the time axis with TRON - II. MeerKAT detects a stellar radio flare from a distant RS CVn candidate}",
      journal = {\mnras},
         year = 2025,
        month = mar,
       volume = {538},
       number = {1},
        pages = {L89-L93},
          doi = {10.1093/mnrasl/slaf015},
archivePrefix = {arXiv},
       eprint = {2501.09489},
 primaryClass = {astro-ph.IM},
       adsurl = {https://ui.adsabs.harvard.edu/abs/2025MNRAS.538L..89S}
}

@ARTICLE{Andersson2022,
       author = {{Andersson}, Alex and {Fender}, Rob P. and {Lintott}, Chris J. and {Williams}, David R.~A. and {Driessen}, Laura N. and {Woudt}, Patrick A. and {van der Horst}, Alexander J. and {Buckley}, David A.~H. and {Motta}, Sara E. and {Rhodes}, Lauren and {Eisner}, Nora L. and {Osten}, Rachel A. and {Vreeswijk}, Paul and {Bloemen}, Steven and {Groot}, Paul J.},
        title = "{Serendipitous discovery of radio flaring behaviour from a nearby M dwarf with MeerKAT}",
      journal = {\mnras},
         year = 2022,
        month = jul,
       volume = {513},
       number = {3},
        pages = {3482-3492},
          doi = {10.1093/mnras/stac1002},
archivePrefix = {arXiv},
       eprint = {2204.03481},
 primaryClass = {astro-ph.SR},
       adsurl = {https://ui.adsabs.harvard.edu/abs/2022MNRAS.513.3482A}
}

@ARTICLE{Shulyak2019,
       author = {{Shulyak}, D. and {Reiners}, A. and {Nagel}, E. and {Tal-Or}, L. and {Caballero}, J.~A. and {Zechmeister}, M. and {B{\'e}jar}, V.~J.~S. and {Cort{\'e}s-Contreras}, M. and {Martin}, E.~L. and {Kaminski}, A. and {Ribas}, I. and {Quirrenbach}, A. and {Amado}, P.~J. and {Anglada-Escud{\'e}}, G. and {Bauer}, F.~F. and {Dreizler}, S. and {Guenther}, E.~W. and {Henning}, T. and {Jeffers}, S.~V. and {K{\"u}rster}, M. and {Lafarga}, M. and {Montes}, D. and {Morales}, J.~C. and {Pedraz}, S.},
        title = "{Magnetic fields in M dwarfs from the CARMENES survey}",
      journal = {\aap},
         year = 2019,
        month = jun,
       volume = {626},
          eid = {A86},
        pages = {A86},
          doi = {10.1051/0004-6361/201935315},
archivePrefix = {arXiv},
       eprint = {1904.12762},
 primaryClass = {astro-ph.SR},
       adsurl = {https://ui.adsabs.harvard.edu/abs/2019A&A...626A..86S}
}

@ARTICLE{Callingham2021,
       author = {{Callingham}, J.~R. and {Vedantham}, H.~K. and {Shimwell}, T.~W. and {Pope}, B.~J.~S. and {Davis}, I.~E. and {Best}, P.~N. and {Hardcastle}, M.~J. and {R{\"o}ttgering}, H.~J.~A. and {Sabater}, J. and {Tasse}, C. and {van Weeren}, R.~J. and {Williams}, W.~L. and {Zarka}, P. and {de Gasperin}, F. and {Drabent}, A.},
        title = "{The population of M dwarfs observed at low radio frequencies}",
      journal = {Nature Astronomy},
         year = 2021,
        month = dec,
       volume = {5},
        pages = {1233-1239},
          doi = {10.1038/s41550-021-01483-0},
archivePrefix = {arXiv},
       eprint = {2110.03713},
 primaryClass = {astro-ph.SR},
       adsurl = {https://ui.adsabs.harvard.edu/abs/2021NatAs...5.1233C}
}

@ARTICLE{Shields2016,
       author = {{Shields}, Aomawa L. and {Ballard}, Sarah and {Johnson}, John Asher},
        title = "{The habitability of planets orbiting M-dwarf stars}",
      journal = {\physrep},
         year = 2016,
        month = dec,
       volume = {663},
        pages = {1},
          doi = {10.1016/j.physrep.2016.10.003},
archivePrefix = {arXiv},
       eprint = {1610.05765},
 primaryClass = {astro-ph.EP},
       adsurl = {https://ui.adsabs.harvard.edu/abs/2016PhR...663....1S}
}

@PROCEEDINGS{Mathys2001,
        title = "{Magnetic Fields Across the Hertzsprung-Russell Diagram}",
    booktitle = {Magnetic Fields Across the Hertzsprung-Russell Diagram},
         year = 2001,
       editor = {{Mathys}, G. and {Solanki}, S.~K. and {Wickramasinghe}, D.~T.},
       series = {Astronomical Society of the Pacific Conference Series},
       volume = {248},
        month = jan,
       adsurl = {https://ui.adsabs.harvard.edu/abs/2001ASPC..248.....M}
}

@ARTICLE{Charbonneau2010,
       author = {{Charbonneau}, Paul},
        title = "{Dynamo Models of the Solar Cycle}",
      journal = {Living Reviews in Solar Physics},
         year = 2010,
        month = dec,
       volume = {7},
       number = {1},
          eid = {3},
        pages = {3},
          doi = {10.12942/lrsp-2010-3},
       adsurl = {https://ui.adsabs.harvard.edu/abs/2010LRSP....7....3C}
}

@ARTICLE{Browning2008,
       author = {{Browning}, Matthew K.},
        title = "{Simulations of Dynamo Action in Fully Convective Stars}",
      journal = {\apj},
         year = 2008,
        month = apr,
       volume = {676},
       number = {2},
        pages = {1262-1280},
          doi = {10.1086/527432},
archivePrefix = {arXiv},
       eprint = {0712.1603},
 primaryClass = {astro-ph},
       adsurl = {https://ui.adsabs.harvard.edu/abs/2008ApJ...676.1262B}
}

@ARTICLE{Dobler2006,
       author = {{Dobler}, Wolfgang and {Stix}, Michael and {Brandenburg}, Axel},
        title = "{Magnetic Field Generation in Fully Convective Rotating Spheres}",
      journal = {\apj},
         year = 2006,
        month = feb,
       volume = {638},
       number = {1},
        pages = {336-347},
          doi = {10.1086/498634},
archivePrefix = {arXiv},
       eprint = {astro-ph/0410645},
 primaryClass = {astro-ph},
       adsurl = {https://ui.adsabs.harvard.edu/abs/2006ApJ...638..336D}
}

@ARTICLE{Chabrier2006,
       author = {{Chabrier}, G. and {K{\"u}ker}, M.},
        title = "{Large-scale {\ensuremath{\alpha}}\^2-dynamo in low-mass stars and brown dwarfs}",
      journal = {\aap},
         year = 2006,
        month = feb,
       volume = {446},
       number = {3},
        pages = {1027-1037},
          doi = {10.1051/0004-6361:20042475},
archivePrefix = {arXiv},
       eprint = {astro-ph/0510075},
 primaryClass = {astro-ph},
       adsurl = {https://ui.adsabs.harvard.edu/abs/2006A&A...446.1027C}
}

@ARTICLE{Parker1955,
       author = {{Parker}, Eugene N.},
        title = "{Hydromagnetic Dynamo Models.}",
      journal = {\apj},
         year = 1955,
        month = sep,
       volume = {122},
        pages = {293},
          doi = {10.1086/146087},
       adsurl = {https://ui.adsabs.harvard.edu/abs/1955ApJ...122..293P}
}

@ARTICLE{Gloudemans2023,
       author = {{Gloudemans}, A.~J. and {Callingham}, J.~R. and {Duncan}, K.~J. and {Saxena}, A. and {Harikane}, Y. and {Hill}, G.~J. and {Zeimann}, G.~R. and {R{\"o}ttgering}, H.~J.~A. and {Hardcastle}, M.~J. and {Pineda}, J.~S. and {Shimwell}, T.~W. and {Smith}, D.~J.~B. and {Wagenveld}, J.~D.},
        title = "{Plausible association of distant late M dwarfs with low-frequency radio emission}",
      journal = {\aap},
         year = 2023,
        month = oct,
       volume = {678},
          eid = {A161},
        pages = {A161},
          doi = {10.1051/0004-6361/202347141},
archivePrefix = {arXiv},
       eprint = {2309.01741},
 primaryClass = {astro-ph.SR},
       adsurl = {https://ui.adsabs.harvard.edu/abs/2023A&A...678A.161G}
}

@ARTICLE{Golay2023,
       author = {{Golay}, Walter W. and {Mutel}, Robert L. and {Lipman}, Dani and {G{\"u}del}, Manuel},
        title = "{A search for thermal gyro-synchrotron emission from hot stellar coronae}",
      journal = {\mnras},
         year = 2023,
        month = jun,
       volume = {522},
       number = {1},
        pages = {1394-1410},
          doi = {10.1093/mnras/stad980},
archivePrefix = {arXiv},
       eprint = {2210.11440},
 primaryClass = {astro-ph.SR},
       adsurl = {https://ui.adsabs.harvard.edu/abs/2023MNRAS.522.1394G}
}

@ARTICLE{Dulk1985,
       author = {{Dulk}, G.~A.},
        title = "{Radio emission from the sun and stars.}",
      journal = {\araa},
         year = 1985,
        month = jan,
       volume = {23},
        pages = {169-224},
          doi = {10.1146/annurev.aa.23.090185.001125},
       adsurl = {https://ui.adsabs.harvard.edu/abs/1985ARA&A..23..169D}
}

@ARTICLE{Gudel2002,
       author = {{G{\"u}del}, Manuel},
        title = "{Stellar Radio Astronomy: Probing Stellar Atmospheres from Protostars to Giants}",
      journal = {\araa},
         year = 2002,
        month = jan,
       volume = {40},
        pages = {217-261},
          doi = {10.1146/annurev.astro.40.060401.093806},
archivePrefix = {arXiv},
       eprint = {astro-ph/0206436},
 primaryClass = {astro-ph},
       adsurl = {https://ui.adsabs.harvard.edu/abs/2002ARA&A..40..217G}
}

@ARTICLE{Lindegren2021,
       author = {{Lindegren}, L. and {Klioner}, S.~A. and {Hern{\'a}ndez}, J. and {Bombrun}, A. and {Ramos-Lerate}, M. and {Steidelm{\"u}ller}, H. and {Bastian}, U. and {Biermann}, M. and {de Torres}, A. and {Gerlach}, E. and {Geyer}, R. and {Hilger}, T. and {Hobbs}, D. and {Lammers}, U. and {McMillan}, P.~J. and {Stephenson}, C.~A. and {Casta{\~n}eda}, J. and {Davidson}, M. and {Fabricius}, C. and {Gracia-Abril}, G. and {Portell}, J. and {Rowell}, N. and {Teyssier}, D. and {Torra}, F. and {Bartolom{\'e}}, S. and {Clotet}, M. and {Garralda}, N. and {Gonz{\'a}lez-Vidal}, J.~J. and {Torra}, J. and {Abbas}, U. and {Altmann}, M. and {Anglada Varela}, E. and {Balaguer-N{\'u}{\~n}ez}, L. and {Balog}, Z. and {Barache}, C. and {Becciani}, U. and {Bernet}, M. and {Bertone}, S. and {Bianchi}, L. and {Bouquillon}, S. and {Brown}, A.~G.~A. and {Bucciarelli}, B. and {Busonero}, D. and {Butkevich}, A.~G. and {Buzzi}, R. and {Cancelliere}, R. and {Carlucci}, T. and {Charlot}, P. and {Cioni}, M.-R.~L. and {Crosta}, M. and {Crowley}, C. and {del Peloso}, E.~F. and {del Pozo}, E. and {Drimmel}, R. and {Esquej}, P. and {Fienga}, A. and {Fraile}, E. and {Gai}, M. and {Garcia-Reinaldos}, M. and {Guerra}, R. and {Hambly}, N.~C. and {Hauser}, M. and {Jan{\ss}en}, K. and {Jordan}, S. and {Kostrzewa-Rutkowska}, Z. and {Lattanzi}, M.~G. and {Liao}, S. and {Licata}, E. and {Lister}, T.~A. and {L{\"o}ffler}, W. and {Marchant}, J.~M. and {Masip}, A. and {Mignard}, F. and {Mints}, A. and {Molina}, D. and {Mora}, A. and {Morbidelli}, R. and {Murphy}, C.~P. and {Pagani}, C. and {Panuzzo}, P. and {Pe{\~n}alosa Esteller}, X. and {Poggio}, E. and {Re Fiorentin}, P. and {Riva}, A. and {Sagrist{\`a} Sell{\'e}s}, A. and {Sanchez Gimenez}, V. and {Sarasso}, M. and {Sciacca}, E. and {Siddiqui}, H.~I. and {Smart}, R.~L. and {Souami}, D. and {Spagna}, A. and {Steele}, I.~A. and {Taris}, F. and {Utrilla}, E. and {van Reeven}, W. and {Vecchiato}, A.},
        title = "{Gaia Early Data Release 3. The astrometric solution}",
      journal = {\aap},
         year = 2021,
        month = may,
       volume = {649},
          eid = {A2},
        pages = {A2},
          doi = {10.1051/0004-6361/202039709},
archivePrefix = {arXiv},
       eprint = {2012.03380},
 primaryClass = {astro-ph.IM},
       adsurl = {https://ui.adsabs.harvard.edu/abs/2021A&A...649A...2L}
}

@ARTICLE{Baraffe2018,
       author = {{Baraffe}, Isabelle and {Chabrier}, Gilles},
        title = "{A closer look at the transition between fully convective and partly radiative low-mass stars}",
      journal = {\aap},
         year = 2018,
        month = nov,
       volume = {619},
          eid = {A177},
        pages = {A177},
          doi = {10.1051/0004-6361/201834062},
archivePrefix = {arXiv},
       eprint = {1809.07274},
 primaryClass = {astro-ph.SR},
       adsurl = {https://ui.adsabs.harvard.edu/abs/2018A&A...619A.177B}
}

@ARTICLE{Reiners2014,
       author = {{Reiners}, A. and {Sch{\"u}ssler}, M. and {Passegger}, V.~M.},
        title = "{Generalized Investigation of the Rotation-Activity Relation: Favoring Rotation Period instead of Rossby Number}",
      journal = {\apj},
         year = 2014,
        month = oct,
       volume = {794},
       number = {2},
          eid = {144},
        pages = {144},
          doi = {10.1088/0004-637X/794/2/144},
archivePrefix = {arXiv},
       eprint = {1408.6175},
 primaryClass = {astro-ph.SR},
       adsurl = {https://ui.adsabs.harvard.edu/abs/2014ApJ...794..144R}
}

@ARTICLE{Kellerman1969,
       author = {{Kellermann}, K.~I. and {Pauliny-Toth}, I.~I.~K.},
        title = "{The Spectra of Opaque Radio Sources}",
      journal = {\apjl},
         year = 1969,
        month = feb,
       volume = {155},
        pages = {L71},
          doi = {10.1086/180305},
       adsurl = {https://ui.adsabs.harvard.edu/abs/1969ApJ...155L..71K}
}

@ARTICLE{Venot2016,
       author = {{Venot}, Olivia and {Rocchetto}, Marco and {Carl}, Shaun and {Roshni Hashim}, Aysha and {Decin}, Leen},
        title = "{Influence of Stellar Flares on the Chemical Composition of Exoplanets and Spectra}",
      journal = {\apj},
         year = 2016,
        month = oct,
       volume = {830},
       number = {2},
          eid = {77},
        pages = {77},
          doi = {10.3847/0004-637X/830/2/77},
archivePrefix = {arXiv},
       eprint = {1607.08147},
 primaryClass = {astro-ph.EP},
       adsurl = {https://ui.adsabs.harvard.edu/abs/2016ApJ...830...77V}
}

@ARTICLE{Nicholls2023,
       author = {{Nicholls}, Harrison and {H{\'e}brard}, Eric and {Venot}, Olivia and {Drummond}, Benjamin and {Evans}, Elise},
        title = "{Temperature-chemistry coupling in the evolution of gas giant atmospheres driven by stellar flares}",
      journal = {\mnras},
         year = 2023,
        month = aug,
       volume = {523},
       number = {4},
        pages = {5681-5702},
          doi = {10.1093/mnras/stad1734},
archivePrefix = {arXiv},
       eprint = {2306.03673},
 primaryClass = {astro-ph.EP},
       adsurl = {https://ui.adsabs.harvard.edu/abs/2023MNRAS.523.5681N}
}

@ARTICLE{Pineda2023,
       author = {{Pineda}, J. Sebastian and {Villadsen}, Jackie},
        title = "{Coherent radio bursts from known M-dwarf planet-host YZ Ceti}",
      journal = {Nature Astronomy},
         year = 2023,
        month = may,
       volume = {7},
        pages = {569-578},
          doi = {10.1038/s41550-023-01914-0},
archivePrefix = {arXiv},
       eprint = {2304.00031},
 primaryClass = {astro-ph.SR},
       adsurl = {https://ui.adsabs.harvard.edu/abs/2023NatAs...7..569P}
}

@ARTICLE{Wu1979,
       author = {{Wu}, C.~S. and {Lee}, L.~C.},
        title = "{A theory of the terrestrial kilometric radiation.}",
      journal = {\apj},
         year = 1979,
        month = jun,
       volume = {230},
        pages = {621-626},
          doi = {10.1086/157120},
       adsurl = {https://ui.adsabs.harvard.edu/abs/1979ApJ...230..621W}
}

@ARTICLE{Mamajeck2013,
       author = {{Pecaut}, Mark J. and {Mamajek}, Eric E.},
        title = "{Intrinsic Colors, Temperatures, and Bolometric Corrections of Pre-main-sequence Stars}",
      journal = {\apjs},
         year = 2013,
        month = sep,
       volume = {208},
       number = {1},
          eid = {9},
        pages = {9},
          doi = {10.1088/0067-0049/208/1/9},
archivePrefix = {arXiv},
       eprint = {1307.2657},
 primaryClass = {astro-ph.SR},
       adsurl = {https://ui.adsabs.harvard.edu/abs/2013ApJS..208....9P}
}

@ARTICLE{Henry2006,
       author = {{Henry}, Todd J. and {Jao}, Wei-Chun and {Subasavage}, John P. and {Beaulieu}, Thomas D. and {Ianna}, Philip A. and {Costa}, Edgardo and {M{\'e}ndez}, Ren{\'e} A.},
        title = "{The Solar Neighborhood. XVII. Parallax Results from the CTIOPI 0.9 m Program: 20 New Members of the RECONS 10 Parsec Sample}",
      journal = {\aj},
         year = 2006,
        month = dec,
       volume = {132},
       number = {6},
        pages = {2360-2371},
          doi = {10.1086/508233},
archivePrefix = {arXiv},
       eprint = {astro-ph/0608230},
 primaryClass = {astro-ph},
       adsurl = {https://ui.adsabs.harvard.edu/abs/2006AJ....132.2360H}
}

@ARTICLE{Madhu2023,
       author = {{Madhusudhan}, Nikku and {Sarkar}, Subhajit and {Constantinou}, Savvas and {Holmberg}, M{\r{a}}ns and {Piette}, Anjali A.~A. and {Moses}, Julianne I.},
        title = "{Carbon-bearing Molecules in a Possible Hycean Atmosphere}",
      journal = {\apjl},
         year = 2023,
        month = oct,
       volume = {956},
       number = {1},
          eid = {L13},
        pages = {L13},
          doi = {10.3847/2041-8213/acf577},
archivePrefix = {arXiv},
       eprint = {2309.05566},
 primaryClass = {astro-ph.EP},
       adsurl = {https://ui.adsabs.harvard.edu/abs/2023ApJ...956L..13M}
}

@ARTICLE{Cloutier2017,
       author = {{Cloutier}, R. and {Astudillo-Defru}, N. and {Doyon}, R. and {Bonfils}, X. and {Almenara}, J. -M. and {Benneke}, B. and {Bouchy}, F. and {Delfosse}, X. and {Ehrenreich}, D. and {Forveille}, T. and {Lovis}, C. and {Mayor}, M. and {Menou}, K. and {Murgas}, F. and {Pepe}, F. and {Rowe}, J. and {Santos}, N.~C. and {Udry}, S. and {W{\"u}nsche}, A.},
        title = "{Characterization of the K2-18 multi-planetary system with HARPS. A habitable zone super-Earth and discovery of a second, warm super-Earth on a non-coplanar orbit}",
      journal = {\aap},
         year = 2017,
        month = dec,
       volume = {608},
          eid = {A35},
        pages = {A35},
          doi = {10.1051/0004-6361/201731558},
archivePrefix = {arXiv},
       eprint = {1707.04292},
 primaryClass = {astro-ph.EP},
       adsurl = {https://ui.adsabs.harvard.edu/abs/2017A&A...608A..35C}
}

@ARTICLE{Kaltnegger2011,
       author = {{Kaltenegger}, Lisa and {Segura}, Ant{\'\i}gona and {Mohanty}, Subhanjoy},
        title = "{Model Spectra of the First Potentially Habitable Super-Earth{\textemdash}Gl581d}",
      journal = {\apj},
         year = 2011,
        month = may,
       volume = {733},
       number = {1},
          eid = {35},
        pages = {35},
          doi = {10.1088/0004-637X/733/1/35},
archivePrefix = {arXiv},
       eprint = {1103.2953},
 primaryClass = {astro-ph.EP},
       adsurl = {https://ui.adsabs.harvard.edu/abs/2011ApJ...733...35K}
}

@ARTICLE{Madhu2021,
       author = {{Madhusudhan}, Nikku and {Piette}, Anjali A.~A. and {Constantinou}, Savvas},
        title = "{Habitability and Biosignatures of Hycean Worlds}",
      journal = {\apj},
         year = 2021,
        month = sep,
       volume = {918},
       number = {1},
          eid = {1},
        pages = {1},
          doi = {10.3847/1538-4357/abfd9c},
archivePrefix = {arXiv},
       eprint = {2108.10888},
 primaryClass = {astro-ph.EP},
       adsurl = {https://ui.adsabs.harvard.edu/abs/2021ApJ...918....1M}
}

@INPROCEEDINGS{McMullin2007,
       author = {{McMullin}, J.~P. and {Waters}, B. and {Schiebel}, D. and {Young}, W. and {Golap}, K.},
        title = "{CASA Architecture and Applications}",
    booktitle = {Astronomical Data Analysis Software and Systems XVI},
         year = 2007,
       editor = {{Shaw}, R.~A. and {Hill}, F. and {Bell}, D.~J.},
       series = {Astronomical Society of the Pacific Conference Series},
       volume = {376},
        month = oct,
        pages = {127},
       adsurl = {https://ui.adsabs.harvard.edu/abs/2007ASPC..376..127M}
}

@ARTICLE{Schweitzer2019,
       author = {{Schweitzer}, A. and {Passegger}, V.~M. and {Cifuentes}, C. and {B{\'e}jar}, V.~J.~S. and {Cort{\'e}s-Contreras}, M. and {Caballero}, J.~A. and {del Burgo}, C. and {Czesla}, S. and {K{\"u}rster}, M. and {Montes}, D. and {Zapatero Osorio}, M.~R. and {Ribas}, I. and {Reiners}, A. and {Quirrenbach}, A. and {Amado}, P.~J. and {Aceituno}, J. and {Anglada-Escud{\'e}}, G. and {Bauer}, F.~F. and {Dreizler}, S. and {Jeffers}, S.~V. and {Guenther}, E.~W. and {Henning}, T. and {Kaminski}, A. and {Lafarga}, M. and {Marfil}, E. and {Morales}, J.~C. and {Schmitt}, J.~H.~M.~M. and {Seifert}, W. and {Solano}, E. and {Tabernero}, H.~M. and {Zechmeister}, M.},
        title = "{The CARMENES search for exoplanets around M dwarfs. Different roads to radii and masses of the target stars}",
      journal = {\aap},
         year = 2019,
        month = may,
       volume = {625},
          eid = {A68},
        pages = {A68},
          doi = {10.1051/0004-6361/201834965},
archivePrefix = {arXiv},
       eprint = {1904.03231},
 primaryClass = {astro-ph.SR},
       adsurl = {https://ui.adsabs.harvard.edu/abs/2019A&A...625A..68S}
}

@ARTICLE{Reid2017,
       author = {{Reid}, Hamish A.~S. and {Kontar}, Eduard P.},
        title = "{Langmuir wave electric fields induced by electron beams in the heliosphere}",
      journal = {\aap},
         year = 2017,
        month = feb,
       volume = {598},
          eid = {A44},
        pages = {A44},
          doi = {10.1051/0004-6361/201629697},
archivePrefix = {arXiv},
       eprint = {1611.07901},
 primaryClass = {astro-ph.SR},
       adsurl = {https://ui.adsabs.harvard.edu/abs/2017A&A...598A..44R}
}

@ARTICLE{Montet2015,
       author = {{Montet}, Benjamin T. and {Morton}, Timothy D. and {Foreman-Mackey}, Daniel and {Johnson}, John Asher and {Hogg}, David W. and {Bowler}, Brendan P. and {Latham}, David W. and {Bieryla}, Allyson and {Mann}, Andrew W.},
        title = "{Stellar and Planetary Properties of K2 Campaign 1 Candidates and Validation of 17 Planets, Including a Planet Receiving Earth-like Insolation}",
      journal = {\apj},
         year = 2015,
        month = aug,
       volume = {809},
       number = {1},
          eid = {25},
        pages = {25},
          doi = {10.1088/0004-637X/809/1/25},
archivePrefix = {arXiv},
       eprint = {1503.07866},
 primaryClass = {astro-ph.EP},
       adsurl = {https://ui.adsabs.harvard.edu/abs/2015ApJ...809...25M}
}

@ARTICLE{Todd2024,
       author = {{Henry}, Todd J. and {Jao}, Wei-Chun},
        title = "{The Character of M Dwarfs}",
      journal = {\araa},
         year = 2024,
        month = sep,
       volume = {62},
       number = {1},
        pages = {593-633},
          doi = {10.1146/annurev-astro-052722-102740},
       adsurl = {https://ui.adsabs.harvard.edu/abs/2024ARA&A..62..593H}
}

@ARTICLE{Driessen2022,
       author = {{Driessen}, L.~N. and {Williams}, D.~R.~A. and {McDonald}, I. and {Stappers}, B.~W. and {Buckley}, D.~A.~H. and {Fender}, R.~P. and {Woudt}, P.~A.},
        title = "{The detection of radio emission from known X-ray flaring star EXO 040830-7134.7}",
      journal = {\mnras},
         year = 2022,
        month = feb,
       volume = {510},
       number = {1},
        pages = {1083-1092},
          doi = {10.1093/mnras/stab3461},
archivePrefix = {arXiv},
       eprint = {2111.13283},
 primaryClass = {astro-ph.SR},
       adsurl = {https://ui.adsabs.harvard.edu/abs/2022MNRAS.510.1083D}
}

@ARTICLE{Burgasser2005,
       author = {{Burgasser}, Adam J. and {Putman}, Mary E.},
        title = "{Quiescent Radio Emission from Southern Late-Type M Dwarfs and a Spectacular Radio Flare from the M8 Dwarf DENIS 1048-3956}",
      journal = {\apj},
         year = 2005,
        month = jun,
       volume = {626},
       number = {1},
        pages = {486-497},
          doi = {10.1086/429788},
archivePrefix = {arXiv},
       eprint = {astro-ph/0502365},
 primaryClass = {astro-ph},
       adsurl = {https://ui.adsabs.harvard.edu/abs/2005ApJ...626..486B}
}

@ARTICLE{Villadsen2019,
       author = {{Villadsen}, Jackie and {Hallinan}, Gregg},
        title = "{Ultra-wideband Detection of 22 Coherent Radio Bursts on M Dwarfs}",
      journal = {\apj},
         year = 2019,
        month = feb,
       volume = {871},
       number = {2},
          eid = {214},
        pages = {214},
          doi = {10.3847/1538-4357/aaf88e},
archivePrefix = {arXiv},
       eprint = {1810.00855},
 primaryClass = {astro-ph.SR},
       adsurl = {https://ui.adsabs.harvard.edu/abs/2019ApJ...871..214V}
}

@ARTICLE{Madhusudhan2025,
       author = {{Madhusudhan}, Nikku and {Constantinou}, Savvas and {Holmberg}, M{\r{a}}ns and {Sarkar}, Subhajit and {Piette}, Anjali A.~A. and {Moses}, Julianne I.},
        title = "{New Constraints on DMS and DMDS in the Atmosphere of K2-18 b from JWST MIRI}",
      journal = {\apjl},
         year = 2025,
        month = apr,
       volume = {983},
       number = {2},
          eid = {L40},
        pages = {L40},
          doi = {10.3847/2041-8213/adc1c8},
archivePrefix = {arXiv},
       eprint = {2504.12267},
 primaryClass = {astro-ph.EP},
       adsurl = {https://ui.adsabs.harvard.edu/abs/2025ApJ...983L..40M}
}

@ARTICLE{Gunther2020,
       author = {{G{\"u}nther}, Maximilian N. and {Zhan}, Zhuchang and {Seager}, Sara and {Rimmer}, Paul B. and {Ranjan}, Sukrit and {Stassun}, Keivan G. and {Oelkers}, Ryan J. and {Daylan}, Tansu and {Newton}, Elisabeth and {Kristiansen}, Martti H. and {Olah}, Katalin and {Gillen}, Edward and {Rappaport}, Saul and {Ricker}, George R. and {Vanderspek}, Roland K. and {Latham}, David W. and {Winn}, Joshua N. and {Jenkins}, Jon M. and {Glidden}, Ana and {Fausnaugh}, Michael and {Levine}, Alan M. and {Dittmann}, Jason A. and {Quinn}, Samuel N. and {Krishnamurthy}, Akshata and {Ting}, Eric B.},
        title = "{Stellar Flares from the First TESS Data Release: Exploring a New Sample of M Dwarfs}",
      journal = {\aj},
         year = 2020,
        month = feb,
       volume = {159},
       number = {2},
          eid = {60},
        pages = {60},
          doi = {10.3847/1538-3881/ab5d3a},
archivePrefix = {arXiv},
       eprint = {1901.00443},
 primaryClass = {astro-ph.EP},
       adsurl = {https://ui.adsabs.harvard.edu/abs/2020AJ....159...60G}
}

@ARTICLE{France2013,
       author = {{France}, Kevin and {Froning}, Cynthia S. and {Linsky}, Jeffrey L. and {Roberge}, Aki and {Stocke}, John T. and {Tian}, Feng and {Bushinsky}, Rachel and {D{\'e}sert}, Jean-Michel and {Mauas}, Pablo and {Vieytes}, Mariela and {Walkowicz}, Lucianne M.},
        title = "{The Ultraviolet Radiation Environment around M dwarf Exoplanet Host Stars}",
      journal = {\apj},
         year = 2013,
        month = feb,
       volume = {763},
       number = {2},
          eid = {149},
        pages = {149},
          doi = {10.1088/0004-637X/763/2/149},
archivePrefix = {arXiv},
       eprint = {1212.4833},
 primaryClass = {astro-ph.EP},
       adsurl = {https://ui.adsabs.harvard.edu/abs/2013ApJ...763..149F}
}

@ARTICLE{Gorman2018,
       author = {{O'Gorman}, E. and {Coughlan}, C.~P. and {Vlemmings}, W. and {Varenius}, E. and {Sirothia}, S. and {Ray}, T.~P. and {Olofsson}, H.},
        title = "{A search for radio emission from exoplanets around evolved stars}",
      journal = {\aap},
         year = 2018,
        month = apr,
       volume = {612},
          eid = {A52},
        pages = {A52},
          doi = {10.1051/0004-6361/201731965},
archivePrefix = {arXiv},
       eprint = {1801.07753},
 primaryClass = {astro-ph.EP},
       adsurl = {https://ui.adsabs.harvard.edu/abs/2018A&A...612A..52O}
}

@article{astropy2013,
    Adsurl = {http://adsabs.harvard.edu/abs/2013A%26A...558A..33A},
    Archiveprefix = {arXiv},
    Author = {{Astropy Collaboration} and {Robitaille}, T.~P. and {Tollerud}, E.~J. and {Greenfield}, P. and {Droettboom}, M. and {Bray}, E. and {Aldcroft}, T. and {Davis}, M. and {Ginsburg}, A. and {Price-Whelan}, A.~M. and {Kerzendorf}, W.~E. and {Conley}, A. and {Crighton}, N. and {Barbary}, K. and {Muna}, D. and {Ferguson}, H. and {Grollier}, F. and {Parikh}, M.~M. and {Nair}, P.~H. and {Unther}, H.~M. and {Deil}, C. and {Woillez}, J. and {Conseil}, S. and {Kramer}, R. and {Turner}, J.~E.~H. and {Singer}, L. and {Fox}, R. and {Weaver}, B.~A. and {Zabalza}, V. and {Edwards}, Z.~I. and {Azalee Bostroem}, K. and {Burke}, D.~J. and {Casey}, A.~R. and {Crawford}, S.~M. and {Dencheva}, N. and {Ely}, J. and {Jenness}, T. and {Labrie}, K. and {Lim}, P.~L. and {Pierfederici}, F. and {Pontzen}, A. and {Ptak}, A. and {Refsdal}, B. and {Servillat}, M. and {Streicher}, O.},
    Doi = {10.1051/0004-6361/201322068},
    Eid = {A33},
    Eprint = {1307.6212},
    Journal = {\aap},
    Month = oct,
    Pages = {A33},
    Primaryclass = {astro-ph.IM},
    Title = {{Astropy: A community Python package for astronomy}},
    Volume = 558,
    Year = 2013
}

@ARTICLE{astropy2018,
       author = {{Astropy Collaboration} and {Price-Whelan}, A.~M. and
         {Sip{\H{o}}cz}, B.~M. and {G{\"u}nther}, H.~M. and {Lim}, P.~L. and
         {Crawford}, S.~M. and {Conseil}, S. and {Shupe}, D.~L. and
         {Craig}, M.~W. and {Dencheva}, N. and {Ginsburg}, A. and {Vand
        erPlas}, J.~T. and {Bradley}, L.~D. and {P{\'e}rez-Su{\'a}rez}, D. and
         {de Val-Borro}, M. and {Aldcroft}, T.~L. and {Cruz}, K.~L. and
         {Robitaille}, T.~P. and {Tollerud}, E.~J. and {Ardelean}, C. and
         {Babej}, T. and {Bach}, Y.~P. and {Bachetti}, M. and {Bakanov}, A.~V. and
         {Bamford}, S.~P. and {Barentsen}, G. and {Barmby}, P. and
         {Baumbach}, A. and {Berry}, K.~L. and {Biscani}, F. and {Boquien}, M. and
         {Bostroem}, K.~A. and {Bouma}, L.~G. and {Brammer}, G.~B. and
         {Bray}, E.~M. and {Breytenbach}, H. and {Buddelmeijer}, H. and
         {Burke}, D.~J. and {Calderone}, G. and {Cano Rodr{\'\i}guez}, J.~L. and
         {Cara}, M. and {Cardoso}, J.~V.~M. and {Cheedella}, S. and {Copin}, Y. and
         {Corrales}, L. and {Crichton}, D. and {D'Avella}, D. and {Deil}, C. and
         {Depagne}, {\'E}. and {Dietrich}, J.~P. and {Donath}, A. and
         {Droettboom}, M. and {Earl}, N. and {Erben}, T. and {Fabbro}, S. and
         {Ferreira}, L.~A. and {Finethy}, T. and {Fox}, R.~T. and
         {Garrison}, L.~H. and {Gibbons}, S.~L.~J. and {Goldstein}, D.~A. and
         {Gommers}, R. and {Greco}, J.~P. and {Greenfield}, P. and
         {Groener}, A.~M. and {Grollier}, F. and {Hagen}, A. and {Hirst}, P. and
         {Homeier}, D. and {Horton}, A.~J. and {Hosseinzadeh}, G. and {Hu}, L. and
         {Hunkeler}, J.~S. and {Ivezi{\'c}}, {\v{Z}}. and {Jain}, A. and
         {Jenness}, T. and {Kanarek}, G. and {Kendrew}, S. and {Kern}, N.~S. and
         {Kerzendorf}, W.~E. and {Khvalko}, A. and {King}, J. and {Kirkby}, D. and
         {Kulkarni}, A.~M. and {Kumar}, A. and {Lee}, A. and {Lenz}, D. and
         {Littlefair}, S.~P. and {Ma}, Z. and {Macleod}, D.~M. and
         {Mastropietro}, M. and {McCully}, C. and {Montagnac}, S. and
         {Morris}, B.~M. and {Mueller}, M. and {Mumford}, S.~J. and {Muna}, D. and
         {Murphy}, N.~A. and {Nelson}, S. and {Nguyen}, G.~H. and
         {Ninan}, J.~P. and {N{\"o}the}, M. and {Ogaz}, S. and {Oh}, S. and
         {Parejko}, J.~K. and {Parley}, N. and {Pascual}, S. and {Patil}, R. and
         {Patil}, A.~A. and {Plunkett}, A.~L. and {Prochaska}, J.~X. and
         {Rastogi}, T. and {Reddy Janga}, V. and {Sabater}, J. and
         {Sakurikar}, P. and {Seifert}, M. and {Sherbert}, L.~E. and
         {Sherwood-Taylor}, H. and {Shih}, A.~Y. and {Sick}, J. and
         {Silbiger}, M.~T. and {Singanamalla}, S. and {Singer}, L.~P. and
         {Sladen}, P.~H. and {Sooley}, K.~A. and {Sornarajah}, S. and
         {Streicher}, O. and {Teuben}, P. and {Thomas}, S.~W. and
         {Tremblay}, G.~R. and {Turner}, J.~E.~H. and {Terr{\'o}n}, V. and
         {van Kerkwijk}, M.~H. and {de la Vega}, A. and {Watkins}, L.~L. and
         {Weaver}, B.~A. and {Whitmore}, J.~B. and {Woillez}, J. and
         {Zabalza}, V. and {Astropy Contributors}},
        title = "{The Astropy Project: Building an Open-science Project and Status of the v2.0 Core Package}",
      journal = {\aj},
         year = 2018,
        month = sep,
       volume = {156},
       number = {3},
          eid = {123},
        pages = {123},
          doi = {10.3847/1538-3881/aabc4f},
archivePrefix = {arXiv},
       eprint = {1801.02634},
 primaryClass = {astro-ph.IM},
       adsurl = {https://ui.adsabs.harvard.edu/abs/2018AJ....156..123A}
}

@INPROCEEDINGS{Berdyugina2009,
       author = {{Berdyugina}, Svetlana V.},
        title = "{Stellar magnetic fields across the H-R diagram: observational evidence}",
    booktitle = {Cosmic Magnetic Fields: From Planets, to Stars and Galaxies},
         year = 2009,
       editor = {{Strassmeier}, Klaus G. and {Kosovichev}, Alexander G. and {Beckman}, John E.},
       series = {IAU Symposium},
       volume = {259},
        month = apr,
        pages = {323-332},
          doi = {10.1017/S1743921309030683},
       adsurl = {https://ui.adsabs.harvard.edu/abs/2009IAUS..259..323B}
}

@ARTICLE{Caramazza2023,
       author = {{Caramazza}, M. and {Stelzer}, B. and {Magaudda}, E. and {Raetz}, St. and {G{\"u}del}, M. and {Orlando}, S. and {Poppenh{\"a}ger}, K.},
        title = "{Complete X-ray census of M dwarfs in the solar neighborhood. I. GJ 745 AB: Coronal-hole stars in the 10 pc sample}",
      journal = {\aap},
         year = 2023,
        month = aug,
       volume = {676},
          eid = {A14},
        pages = {A14},
          doi = {10.1051/0004-6361/202346470},
archivePrefix = {arXiv},
       eprint = {2305.14971},
 primaryClass = {astro-ph.SR},
       adsurl = {https://ui.adsabs.harvard.edu/abs/2023A&A...676A..14C}
}

@ARTICLE{dosSantos2020,
       author = {{dos Santos}, Leonardo A. and {Ehrenreich}, David and {Bourrier}, Vincent and {Astudillo-Defru}, Nicola and {Bonfils}, Xavier and {Forget}, Fran{\c{c}}ois and {Lovis}, Christophe and {Pepe}, Francesco and {Udry}, St{\'e}phane},
        title = "{The high-energy environment and atmospheric escape of the mini-Neptune K2-18 b}",
      journal = {\aap},
         year = 2020,
        month = feb,
       volume = {634},
          eid = {L4},
        pages = {L4},
          doi = {10.1051/0004-6361/201937327},
archivePrefix = {arXiv},
       eprint = {2001.04532},
 primaryClass = {astro-ph.EP},
       adsurl = {https://ui.adsabs.harvard.edu/abs/2020A&A...634L...4D}
}






\bsp	
\label{lastpage}
\end{document}